\documentclass[final,twocolumn]{elsarticle}

\usepackage{lineno}
\usepackage{hyperref}
\modulolinenumbers[5]

\usepackage{graphicx}
\usepackage{amsmath}
\usepackage{amssymb}
\usepackage{latexsym}
\usepackage{xcolor}
\usepackage{bm}

\renewcommand *\arraystretch{1.5}

\journal{Nuclear Instruments and Methods in Physics Research Section A}

\begin{document}

\begin{frontmatter}

\title{The Baghdad Atlas: A relational database of inelastic neutron-scattering $(n,n'\gamma)$ data}


\author[UCB]{A.~M.~Hurst\corref{coresp}}
\cortext[coresp]{Corresponding author.}
\ead{amhurst@berkeley.edu}
\ead[URL]{https://nucleardata.berkeley.edu/atlas}
\author[UCB,LBNL]{L.~A.~Bernstein}
\author[LANL]{T.~Kawano}
\author[NNL]{A.~M.~Lewis}
\author[UCB]{K.~Song}

\address[UCB]{Department of Nuclear Engineering, University of California, Berkeley, California 94720, USA}
\address[LBNL]{Lawrence Berkeley National Laboratory, Berkeley, California 94720, USA}
\address[LANL]{Theoretical Division, Los Alamos National Laboratory, Los Alamos, New Mexico 87545, USA}
\address[NNL]{Naval Nuclear Laboratory, Schenectady, New York 12301, USA}

\begin{abstract}

  A relational database has been developed based on the original ($n,n'\gamma$) work carried out by A.~M.~Demidov \textit{et al}., at the Nuclear Research Institute in Baghdad, Iraq [\textit{``Atlas of Gamma-Ray Spectra from the Inelastic Scattering of Reactor Fast Neutrons''}, Nuclear Research Institute, Baghdad, Iraq (Moscow, Atomizdat 1978)] for 105 independent measurements comprising 76 elemental samples of natural composition and 29 isotopically-enriched samples.  The information from this Atlas includes: $\gamma$-ray energies and relative intensities; nuclide and level data corresponding to the residual nucleus and meta data associated with the target sample that allows for the extraction of the flux-weighted ($n,n'\gamma$) cross sections for a given transition relative to a defined value.  The optimized angular-distribution-corrected fast-neutron flux-weighted partial $\gamma$-ray cross section for the production of the 846.8-keV $2^{+}_{1} \rightarrow 0^{+}_{\rm gs}$ $\gamma$-ray transition in $^{56}$Fe, determined to be $\langle \sigma_{\gamma} \rangle = 143(29)$~mb, is used for this purpose.  However, different values for the adopted cross section can be readily implemented to accommodate user preference based on revised determinations of this quantity.  The Atlas ($n,n'\gamma$) data has been compiled into a series of CSV-style ASCII data sets and a suite of Python scripts have been developed to build and install the database locally.  The database can then be accessed directly through the SQLite engine, or using alternative methods such as the Jupyter Notebook Python-browser interface. Several examples exploiting different interaction methodologies are distributed with the complete software package.

\end{abstract}



\end{frontmatter}


\section{Introduction}

Inelastic neutron scattering is a dominant energy-loss mechanism for fast neutrons in heavy ($A>12$) nuclei and produces unique $\gamma$-ray signatures of the material which the neutrons are incident upon.  As such, a good knowledge of it is required for virtually all branches of applied nuclear science ranging from shielding calculations to the design of advanced nuclear-energy systems to international security and counter proliferation.  The need for improved neutron-scattering data was explicitly stated in a number of recent nuclear data workshops, including the white papers from the Nuclear Data Needs and Capabilities for Applications Workshop \cite{NDNCA:2015}, the Nuclear Data Roadmapping Enhancement Workshop in 2018 \cite{NDREW:2018} and the Workshop for Applied Nuclear Data Activities in 2019 \cite{WANDA:2019}.  In addition to its utility for nuclear applications, ($n,n'\gamma$) data provides unique insight into off-yrast nuclear structure due to the non-selective nature of the reaction (which can include a significant compound component) and the wide range of angular momentum states accessible to fast neutrons.

Angle-differential ($n,n'$) data is challenging to measure due to the difficulties involved in measuring the neutron energy and large backgrounds from elastic scattering.  An alternate approach to determining ($n,n'\gamma$) cross sections involves measuring the prompt $\gamma$ rays emitted from the excited states populated via inelastic scattering.  While these measurements lack the angle-differential information that provides a useful testing ground for neutron transport methods, they can provide an important integral constraint to the nuclear-reaction evaluation process and can be used to improve modeling for non-destructive assay of materials using active-neutron interrogation.

Unfortunately, there are no modern compilations of inelastic-scattering $\gamma$-ray production cross sections.  This is in part due to a fundamental lack of data, but also to the fact that since ($n,n'\gamma$) includes both discrete $\gamma$-ray transitions and cross section data it does not fit well into either of the two main compilation databases: Experimental Nuclear Reaction Data (EXFOR) \cite{id:EXFOR, zerkin:18} for reactions, and Experimental Unevaluated Nuclear Data List (XUNDL) \cite{id:XUNDL} for structure.  The \textit{``Atlas of Gamma-rays from the Inelastic Scattering of Reactor Fast Neutrons''} published by A.~M.~Demidov \textit{et al}., \cite{id:ATLAS} is one of the most comprehensive compilations of data on ($n,n'\gamma$) in existence containing 7375 $\gamma$ rays (of these, 6870 represent firm assignments, while 505 are tentative) from 76 natural and 29 isotopically-enriched targets, measured at the IRT-M Reactor at the Nuclear Research Institute, near Baghdad, Iraq.  Until recently this information has only been available in its [out-of-print] book format. To enhance the utility of this Atlas and increase its accessibility to the international community, we have compiled the data into a set of CSV-style ASCII tables and developed software to build the corresponding SQLite (structured query language) database locally.  The software package is disseminated online through the Berkeley Nuclear Data Group \cite{id:BNDG} and the National Nuclear Data Center (NNDC) \cite{id:NNDC}. This paper describes how the data were originally obtained along with query-based methods for extraction of absolute partial $\gamma$-ray production cross sections from the reported \cite{id:ATLAS} relative-intensity data.

\section{Inelastic-scattering ($n,n'\gamma$) data}

The data in the ``Baghdad Atlas'' \cite{id:ATLAS} comes from a single lithium-drifted germanium [Ge(Li)] $\gamma$-ray detector oriented perpendicular to a filtered ``fast'' neutron beam line at the IRT-M reactor, formerly located at the Al-Tuwaitha Nuclear Research Institute outside of Baghdad, Iraq.  While the details of the neutron spectrum are not completely known, the experimental setup was designed to minimize the presence of low-energy neutrons. Their success is evident in that there are only 30 transitions throughout the entire Atlas known to arise from radiative-capture ($n, \gamma$) reactions.  However, it is possible that some of these ($n, \gamma$) signatures are induced by resonance neutrons rather than thermal capture.  In this paper we only provide sufficient details of the neutron flux and and spectroscopy measurements necessary to determine cross sections on an absolute scale for all data contained in the Atlas.  The experimental setup used at the Baghdad reactor is described in greater detail in Refs. \cite{ahmed:74A,ahmed:74B}.

\begin{figure}[ht]
  \includegraphics[angle=0,width=\linewidth]{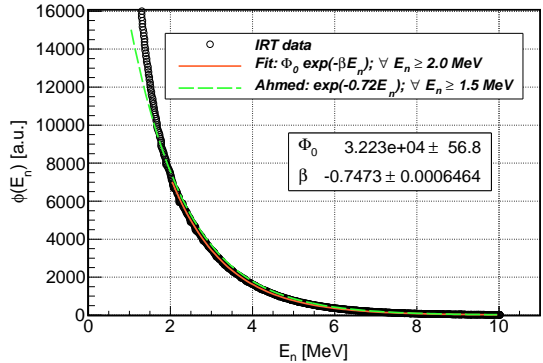}
  \caption{\label{fig:1} The IRT-M Baghdad reactor neutron flux in arbitrary units [a.u.] shown as a function of neutron energy. The data are taken from Ref.~\cite{id:ATLAS} and correspond to the fast-neutron spectrum of the water-cooled water-moderated reactor after filtration of the beam through a 10.1-cm layer comprising lead (9 cm), boron carbide (1 cm) and cadmium (1 mm). The fit (solid-orange curve) corresponds to an exponential of the form $\exp(-\beta E_{n})$, where $\beta = 0.7473$ and $\phi_{0} = 32230$ are determined from a fit to the data in the region $E_{n} \geq 2$~MeV. An exponential of the form $\exp(-0.72 E_{n})$ \cite{ahmed:74A} is also drawn for comparison (dashed-green curve); to obtain this function we fixed the value of $\beta$ to 0.72 and $\phi_{0}$ was treated as a free parameter determined in an independent fit to the region $E_{n} \geq 1.5$~MeV.}
\end{figure}

\subsection{Neutron flux\label{sect:2.1}}

The neutron spectrum is characterized as having a monotonically-decreasing flux ($\phi$) with increasing neutron energy ($E_{n}$), ranging from approximately 0.5 $-$ 10 MeV as shown in Fig.~\ref{fig:1}, where the measured IRT-M data is taken from Refs.~\cite{id:ATLAS, ahmed:74A}.  To a first approximation, the authors of the Atlas suggest the reactor neutron spectrum at $E_{n} > 1.0$~MeV falls off according to the exponential-attenuation law of the form:
\begin{equation}
  \phi(E_{n}) \propto \exp(-\beta E_{n}),
  \label{eq:1}
\end{equation}
where $\beta = 0.7$ \cite{id:ATLAS}.  However, in a previous interpretation of the same measured spectrum, the authors suggest the flux may be approximated in accordance with $\exp(-0.72E_{n})$: $\forall~E_{n} > 1.5$~MeV \cite{ahmed:74A}.  In earlier work still, it is noted that this exponent may, in fact, vary from approximately $\beta \approx 0.65 - 0.75$ \cite{nichol:72} depending on the reactor type.  However, the relative $\gamma$-ray intensities corresponding to transitions from levels above 0.5~MeV are not expected to be strongly affected by different values of $\beta$ \cite{id:ATLAS}.  Because the observed ($n, n'\gamma$) spectra from reactor fast neutrons are largely influenced by the increasing nuclear level density with increasing atomic mass $A$ (away from closed shells), the $\gamma$-ray intensity data presented in the Baghdad Atlas can be expected to be universal for $^{235}$U-fission-based neutron sources \cite{id:ATLAS}.

In our attempts to fit the IRT-M reactor flux data from Ref.~\cite{id:ATLAS} with a single exponential of the form given by Eq.~(\ref{eq:1}), we have determined a value of $\beta = 0.7473$ corresponding to the energy region $E_{n} > 2.0$ MeV, as shown by the fit in Fig.~\ref{fig:1}. Our value of $\beta$ is appreciably larger (approximately 3.8 $-$ 6.8\%) than the values reported in Refs.~\cite{id:ATLAS,ahmed:74A}, but falls within the expected range defined in Ref.~\cite{nichol:72}.  At lower neutron energies the authors suggest that a more complicated function is needed to describe the behaviour of the neutron flux. This is evident from Fig.~\ref{fig:1} where simple exponentials (assuming values of $\beta = 0.72$ \cite{ahmed:74A} and $\beta = 0.7473$) are inadequate at reproducing the reported flux. 

\begin{figure}[t]
  \includegraphics[angle=0,width=\linewidth]{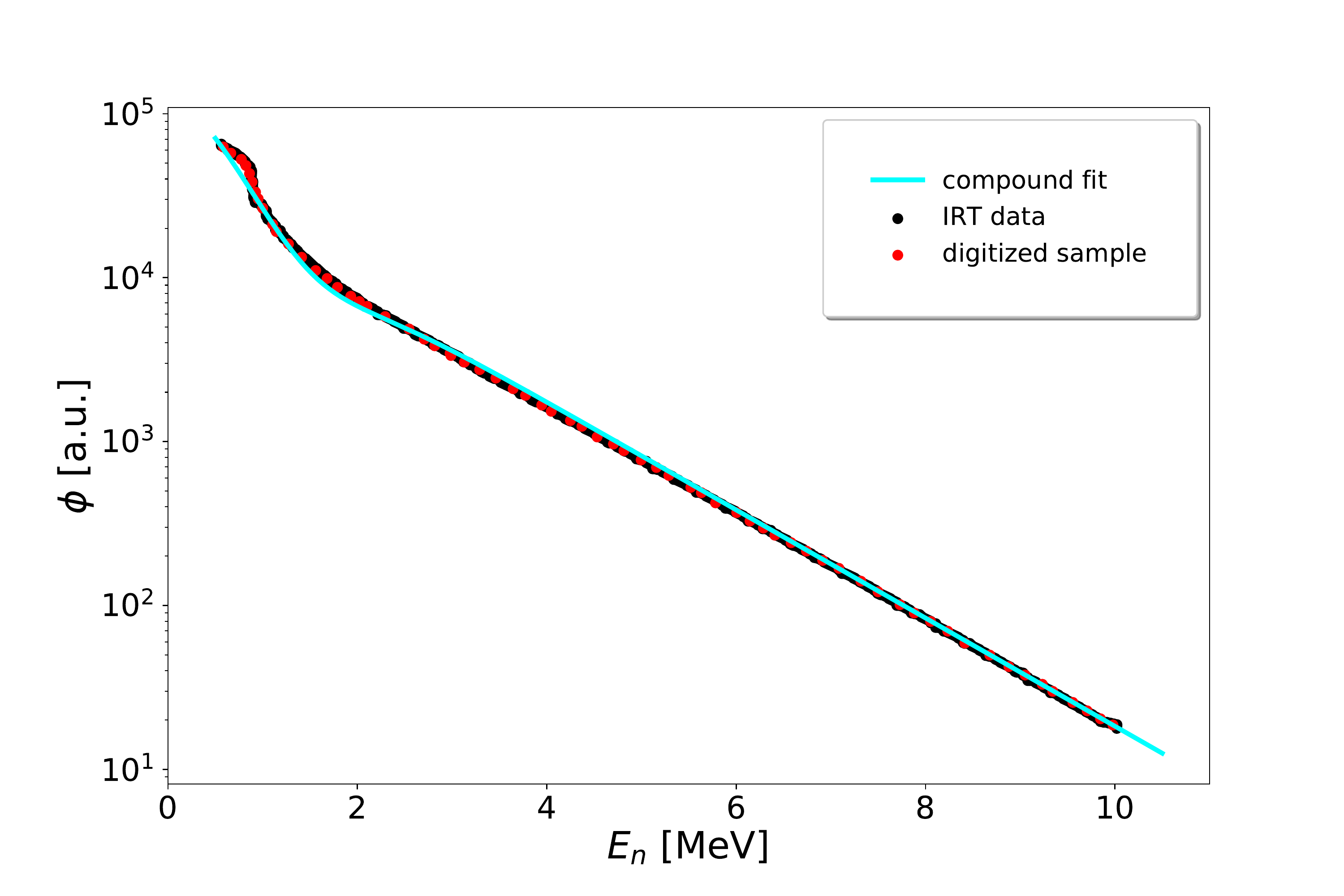}
  \caption{\label{fig:compound_func} The black circles correspond to the IRT-M neutron-flux spectrum digitized from Ref.~\cite{id:ATLAS}, while the red circles are a representative sample of those data points used for fitting purposes.  The resulting compound fit, comprising a low-energy Maxwellian component and a high-energy exponential component as described by Eq.~(\ref{eq:phi_overall}), is illustrated by the solid line.}
\end{figure}

Accordingly, we have attempted to model the flux in the two different regions of the neutron spectrum.  In the \textit{low-energy} region ($E_{n} < 1.5$~MeV) we have considered (i) a Maxwellian distribution of the form
\begin{equation}
  \phi_{1}(E_{n}) = 2A_{1} \sqrt{\left( \frac{E_{n}}{\pi kT^{3}} \right)} \exp\left({\frac{-E_{n}}{kT}}\right),
  \label{eq:phi1}
\end{equation}
where $A_{1}$ [arbitrary units] and $kT$ [MeV] are adjustable parameters optimized from the fit to the low-energy data
.  For the \textit{high-energy} region ($E_{n} \geq 1.5$~MeV) we have used a simple exponential
\begin{equation}
  \phi_{2}(E_{n}) = A_{2} \exp(-\beta E_{n}),
  \label{eq:phi2}
\end{equation}
where $A_{2}$ [arbitrary units] and $\beta$ are adjustable parameters optimized from the fit to the high-energy data.  From the initial fits to the two separate energy regions, we have then modified the $\beta$ parameter further and introduced a smoothing factor, described by an infinitely smooth hyperbolic tangent function, to optimize the fit to both regions.  Thus, the overall fit to the data is given by
\begin{eqnarray}
  \nonumber
  \phi(E_{n}) &=& \phi_{1}(E_{n}) + \left[ \frac{1 + \tanh[K(E_{n} - 1.5)]}{2} \right] \\
  &\times& (\phi_{2}(E_{n}) - \phi_{1}(E_{n})),
  \label{eq:phi_overall}
\end{eqnarray}
where $K \in \mathcal{R}^{+}$ and we have set $K=1.0$ to produce the fit shown in Fig.~\ref{fig:compound_func}.  Equation~(\ref{eq:phi_overall}) implies that for $E_{n} \ll 1.5$~MeV, $\tanh(E_{n}) \rightarrow -1$ and therefore $\phi(E_{n}) \rightarrow \phi_{1}(E_{n})$, whereas for $E_{n} \gg 1.5$~MeV, $\tanh(E_{n}) \rightarrow +1$ and therefore $\phi(E_{n}) \rightarrow \phi_{2}(E_{n})$.  The parametrizations optimized from this fitting procedure are $A_{1} = 84,500$ and $kT = 0.39$~MeV in the Maxwellian region, and $A_{2} = 36,600$ and $\beta = 0.76$ in the exponential tail.  The high-energy component of the neutron-flux spectrum is approximately asymptotic with the suggested parametrizations in the earlier works of Refs.~\cite{id:ATLAS,ahmed:74A} as $E_{n} \rightarrow \infty$.  However, given that we have constrained the Mawellian component of the fit to $E_{n}<1.5$~MeV, our fitted $kT$ value is significantly different to that which is expected for the Maxwellian form of the fission spectrum for $^{235}$U, $kT = 1.29$~MeV \cite{fluence:IAEA}.

\subsection{$\gamma$-ray spectrometry\label{sect:2.2}}

All of the data in the Atlas were acquired under identical conditions using the same experimental configuration with a single Ge(Li) $\gamma$-ray detector oriented at $90^{\circ}$ to the neutron beam line, with samples positioned at $60^{\circ}$ relative to the incident neutron beam in a $25 \times 25$~mm$^{2}$ holder \cite{ahmed:74A}.  A beam-spot diameter of $\sim$ 30 mm was achieved at the sample position \cite{id:ATLAS} and spectra were collected for neutron irradiation periods ranging from 3 $-$ 44~h \cite{id:ATLAS}.  The irradiated samples varied in mass from 1~g (Eu$_{2}$O$_{3}$) to 125~g (Hg).  Unfortunately, however, certain information regarding the samples is lacking from the Atlas, such as the density, thickness, and sample-size relative to the beam spot.  For measurements of most elements and enriched isotopes, a 30~cm$^{3}$ Ge(Li) detector with an energy resolution of 3.8 keV at 1.2 MeV \cite{id:ATLAS,ahmed:74A} was used; it should be noted, however, that the resolution of this detector deteriorated to 8~keV during the course of the measurements because of neutron damage, as mentioned in Ref.~\cite{id:ATLAS}.  A 40~cm$^{3}$ Ge(Li) detector with an energy resolution of 2 keV at 1.2 MeV \cite{id:ATLAS} was also used for the elemental-sample measurements of chlorine, scandium, bromine, lutetium, osmium, and iridium, along with the isotopes of tellurium.

For each measurement, the Ge(Li) detector was isolated from the fast-neutron flux by surrounding it with shielding materials comprised of a 5-cm thick iron plate, an 8-cm layer of paraffin mixed with boron carbide (B$_{4}$C), and a 10-cm thick lead plate. A 20-mm diameter by 110-mm length LiH layer with a density of 0.53 g/cm$^{3}$ \cite{ahmed:74A} was also placed into a collimating channel of the shielding material to filter fast neutrons scattered by the sample itself and to reduce neutron activation of the Ge(Li) detector.

Energy and efficiency calibrations of the Ge(Li) detectors were performed using a variety of standard radioactive ($^{75}$Se, $^{182}$Ta, $^{110}$Ag, $^{72}$Ga, $^{140}$La, $^{24}$Na, and $^{134}$Cs) and reaction [$^{28}$Si($n, \gamma$)$^{29}$Si] sources for energies below and above 3 MeV, respectively.  All $\gamma$-ray intensities ($I_{\gamma}$) were measured at $90^{\circ}$ relative to the neutron beam line and were corrected for self-absorption effects in the sample.  This configuration also helps reduce the effect of Doppler broadening in the $\gamma$ spectra \cite{id:ATLAS}.  The measured $\gamma$-ray energies ($E_{\gamma}$) are reported in the Baghdad Atlas \cite{id:ATLAS} together with uncertainties due to calibration and non-linearity of the spectroscopic tract for $E_{\gamma}$.  The associated uncertainties for the $I_{\gamma}$ measurements listed in the Atlas include statistical contributions (with 95\% probability) together with uncertainties due to detector-efficiency determination and $\gamma$-ray self absorption.

The $\gamma$-ray transition intensities for all elements and enriched isotopes reported in the Atlas are presented in comparison to the 846.8-keV transition in $^{56}$Fe \cite{huo:11}, corresponding to the $2^{+}_{1} \rightarrow 0^{+}_{\rm gs}$ transition observed in the $^{56}$Fe($n, n'\gamma$) reaction. The relative intensity of the 846.8-keV line is defined to be 100\% \cite{id:ATLAS}.  This provides a unique normalization reference point for all other elements (or enriched isotopes) since the intensity of a reliable $\gamma$-ray transition belonging to each element (enriched isotope) was determined relative to the $^{56}$Fe 846.8-keV $\gamma$ line.  The intensities for all other $\gamma$ lines from each respective elemental [enriched isotope] ($n, n'\gamma$) measurement were reported relative to the selected elemental (enriched-isotope) normalization $\gamma$ ray; these normalization $\gamma$-ray lines, together with their $^{56}$Fe-relative-intensity values, are tabulated in [Table 2 of] the Baghdad Atlas \cite{id:ATLAS}.

\section{\label{sect:ang} $\gamma$-ray angular distributions}

For isotropic-distributions, i.e., complete $4 \pi$ solid-angle coverage, the angular-distribution correction factor $W(\theta) = 1$ and no correction to the angular distribution is required.  Other distributions, where $W(\theta) \neq 1$, have some degree of anisotropy that needs to be accounted for.  The angular distribution correction to the inelastic-scattering cross section may be deduced according to the following expression:
\begin{equation}
  \sigma_{\rm inl} = \frac{d\sigma(\theta)}{d \Omega} \frac{4 \pi}{W(\Omega)},
  \label{eq:wcorr1}
\end{equation}
where $\sigma_{\rm inl}$ is the angle-integrated inelastic-scattering cross section, $\sigma(\theta)$ is the inelastic-scattering cross section measured at angle $\theta$, and $\Omega$ is the solid-angle subtended by the detector.  Thus, to find $\sigma(\theta)$:
\begin{eqnarray}
  \nonumber
  \sigma_{\rm inl} \int\limits_{0}^{4\pi} d\Omega &=& \frac{4 \pi}{W(\theta)} \int\limits_{0}^{\sigma(\theta)} d\sigma(\theta)\\
  \sigma(\theta) &=& \sigma_{\rm inl} W(\theta).
  \label{eq:wcorr2}
\end{eqnarray}

According to the prescription outlined by Yamazaki \cite{yamazaki:67}, the angular distribution function for a transition from $J_{i} \rightarrow J_{f}$ may be adequately expressed as
\begin{equation}
  W(\theta) = 1 + A_{2}P_{2}(\cos\theta) + A_{4}P_{4}(\cos\theta),
  \label{eq:wcorr3}
\end{equation}
where $A_{k}$ are the anisotropy coefficients \cite{yamazaki:67} and $P_{k}$ are the Legendre polynomials of order $k$.  The first three \textit{even} polynomial terms may be derived as
\begin{eqnarray}
  \nonumber
  P_{0}(\cos\theta) &=& 1 \\
  \nonumber
  P_{2}(\cos\theta) &=& \frac{1}{2}(3 \cos^{2} \theta - 1) \\
  P_{4}(\cos\theta) &=& \frac{1}{8}(35 \cos^{4} \theta - 30 \cos^{2} \theta + 3).
  \label{eq:wcorr4}
\end{eqnarray}

Equation~(\ref{eq:wcorr3}) assumes complete nuclear alignment of the state undergoing the transition, i.e., equal relative populations of the magnetic substates specified by the population parameter $P_{m}(J) = P_{-m}(J)$, whereupon the anisotropy coefficients$-$tabulated in Ref.~\cite{yamazaki:67}$-$may be written as
\begin{eqnarray}
  \nonumber
  &A_{k}(J_{i} L_{1} L_{2} J_{f})& = \frac{\rho_{k}(J)}{1+\delta_{\gamma}^{2}} [F_{k}(J_{f}L_{1}L_{1}J_{i}) \\
  \nonumber
  &+& 2\delta_{\gamma}F_{k}(J_{f}L_{1}L_{2}J_{i}) \\
  &+& \delta_{\gamma}^{2}F_{k}(J_{f}L_{2}L_{2}J_{i})],
  \label{eq:wcorr5}
\end{eqnarray}
where $\rho_{k}(J)$ is the statistical population tensor given by
\begin{eqnarray}
    \label{eq:wcorr6}
  \rho_{k}(J) &=& \sqrt{2J+1} \\
  \nonumber
  &\times& \sum\limits_{m=-J}^{J} (-1)^{J-m} \langle J m J -m| k 0 \rangle P_{m}(J),
\end{eqnarray}
and the $F_{k}$ distribution coefficients are also tabulated in Ref.~\cite{yamazaki:67} and take the explicit form
\begin{eqnarray}
  \label{eq:wcorr10}
  \nonumber
  F_{k}(J_{f}L_{1}L_{2}J_{i}) &=& (-1)^{J_{f}-J_{i}-1}\\
  \nonumber
  &\times& \sqrt{(2L_{1}+1)(2L_{2}+1)}\\
  \nonumber
  &\times& \sqrt{(2J_{i}+1)} \langle L_{1} 1 L_{2} -1|k 0 \rangle\\
  &\times& W(J_{i}J_{i}L_{1}L_{2}; kJ_{f}),
\end{eqnarray}
where $W(J_{i}J_{i}L_{1}L_{2}; kJ_{f})$ is a Racah recoupling coefficient ($6j$ symbol) \cite{B&M:I}.  For cases where the transition proceeds via a pure stretched $E2$ quadrupole, $L_{1} = L_{2} = 2$, and the interference terms with the $\gamma$-ray multipole-mixing ratio $\delta_{\gamma}$ in Eq.~\ref{eq:wcorr5} vanish, thus reducing the anisotropy coefficient to
\begin{equation}
  A_{k}(J_{i} L_{1} L_{2} J_{f}) = \rho_{k}(J) F_{k}(J_{f}L_{1}L_{1}J_{i}).
  \label{eq:wcorr7}
\end{equation}
For oriented states, $\rho_{k}(J)$ is difficult to solve due to unequal relative populations of the magnetic substates.  However, for a completely aligned state we may replace $\rho_{k}(J)$ in Eq.~(\ref{eq:wcorr6}) with $B_{k}(J)$, such that the statistical tensor may be represented as
\begin{equation}
  \rho_{k}(J) \equiv B_{k}(J) = (-1)^{J} \sqrt{2J+1} \langle J 0 J 0 | k 0 \rangle,
  \label{eq:wcorr8}
\end{equation}
for integral spins, and as
\begin{equation}
  \rho_{k}(J) \equiv B_{k}(J) = (-1)^{J-\frac{1}{2}} \sqrt{2J+1} \langle J \frac{1}{2} J -\frac{1}{2} | k 0 \rangle,
  \label{eq:wcorr8b}
\end{equation}
for half-integral spins, and thus,
\begin{equation}
  A_{k}(J_{i}L_{1}L_{2}J_{f}) = B_{k}(J) F_{k}(J_{f}L_{1}L_{1}J_{i}).
  \label{eq:wcorr9}
\end{equation}

\subsection{\label{sect:ang-s1} Theoretical angular-distribution correction for the $2^{+}_{1} \rightarrow 0^{+}_{\rm gs}$ $\gamma$ ray in $^{56}$Fe\protect\\}

\begin{figure}[t]
  \includegraphics[width=\linewidth, angle=0]{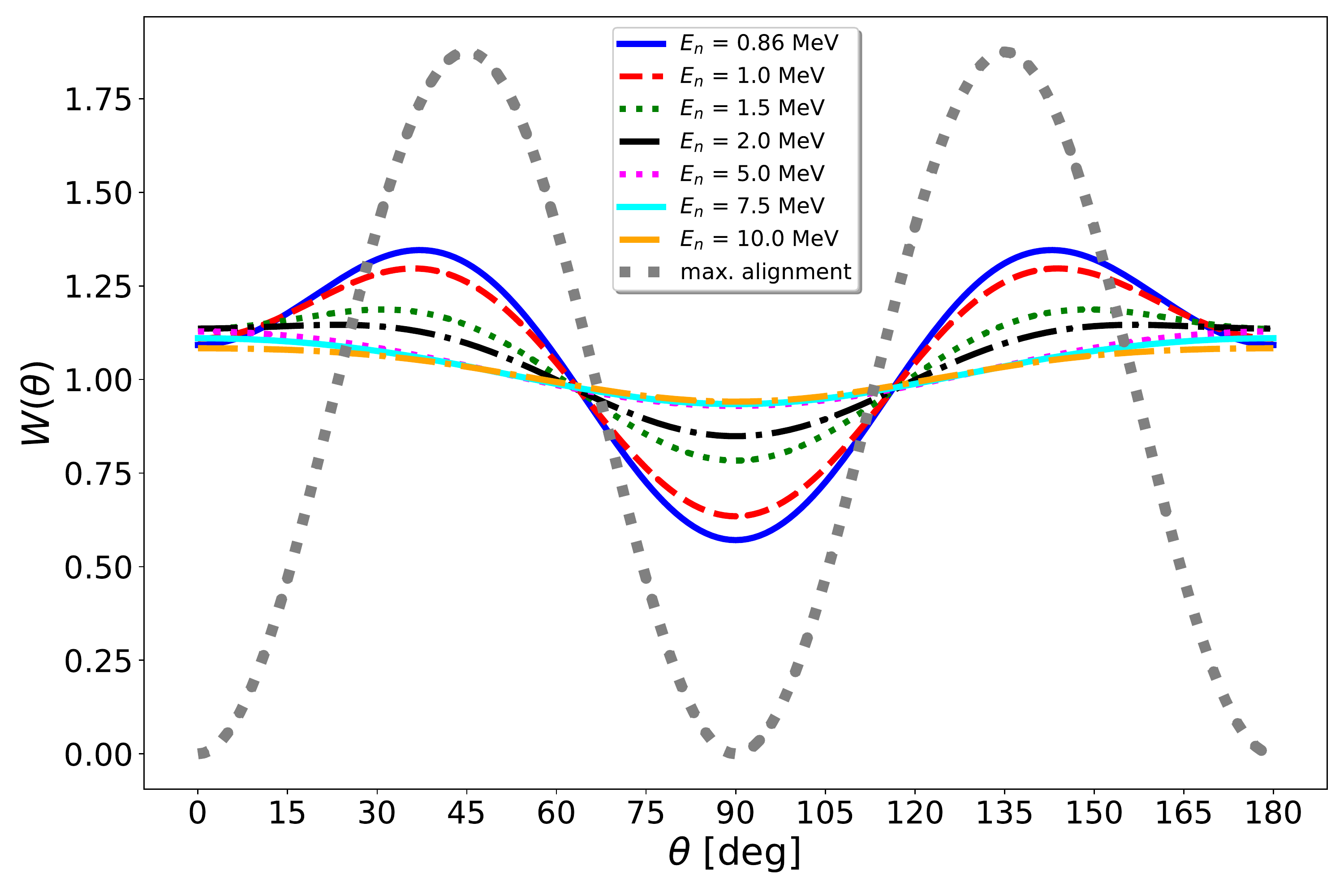}
  \caption{\label{fig:W_calc} Angular distribution functions plotted for the 846.8-keV $\gamma$-ray transition in $^{56}$Fe.  The theoretical distribution is plotted assuming maximum alignment of the initial $J=2$ state in $^{56}$Fe (dotted gray curve).  Experimental $a_{2}$ and $a_{4}$ coefficients \cite{savin:00} at the specified incident-neutron energy $E_{n}$ indicated on the plot have been used to obtain the corresponding experimental distributions.} 
\end{figure}

The 846.8-keV $\gamma$-ray transition in $^{56}$Fe (the normalization transition of the Atlas) proceeds via a pure stretched $E2$ quadrupole ($L_{1} = L_{2} = 2$) from an initial state $J_{i} = 2$ to a final state $J_{f} = 0$.  Assuming complete nuclear alignment, the population tensors for $k=2$ and $k=4$ may be evaluated in accordance with Eq.~(\ref{eq:wcorr8}):
\begin{eqnarray}
  \nonumber
  B_{2}(2) &=& \sqrt{5} \langle 2 0 2 0 | 2 0 \rangle\\
  \nonumber
  B_{4}(2) &=& \sqrt{5} \langle 2 0 2 0 | 4 0 \rangle.
  \label{eq:wtheo1}
\end{eqnarray}
Using the angular momentum coupling coefficient calculator distributed with this project, the above Clebsch-Gordan coefficients are evaluated as $\langle 2 0 2 0 | 2 0 \rangle=-\sqrt{\frac{2}{7}}$ and $\langle 2 0 2 0 | 4 0 \rangle=3\sqrt{\frac{2}{35}}$, thus, we find $B_{2}(2) = -1.19523$ and $B_{4}(2) = 1.60357$.  These computations are in exact agreement with the corresponding $B_{k}$ values presented by Yamazaki in Table~I of Ref.~\cite{yamazaki:67}.  The $F_{k}$ distribution coefficients can then be evaluated using Eq.~(\ref{eq:wcorr10}) where the Racah recoupling coefficient is represented by a $6j$ symbol of the form
\begin{eqnarray}
  \left\{ \begin{array}{ccc}
      J_{i} & L & k \\
      L & J_{i} & J_{f} \end{array} \right\} &=& (-1)^{J_{i} + L + k} \\
  &\times& \frac{1}{\sqrt{2J_{i}+1}} \frac{1}{\sqrt{2L+1}},\nonumber
\label{eq:theo2}
\end{eqnarray}
given that one of the angular momenta, in this case $J_{f}$, vanishes \cite{B&M:I}.  Thus, for both $k=2$ and $k=4$, we find
\begin{equation}
  \nonumber
  \left\{ \begin{array}{ccc}
    2 & 2 & k \\
    2 & 2 & 0 \end{array} \right\} = \frac{1}{5},
  \label{eq:wtheo3}
\end{equation}
which upon substitution for $W(2222; k0)$ in Eq.~(\ref{eq:wcorr10}) gives
\begin{eqnarray}
  \nonumber
  F_{2}(2220) &=& -\sqrt{5} \langle 2 1 2 -1 | 2 0 \rangle \\
  \nonumber
  F_{4}(2220) &=& -\sqrt{5} \langle 2 1 2 -1 | 4 0 \rangle,
  \label{eq:wtheo4}
\end{eqnarray}
yielding evaluated coefficients $F_{2}=-0.59761$ and $F_{4}=-1.06904$.  Again, these computations are in exact correspondence with the $F_{k}$ values presented by Yamazaki in Table~II of Ref.~\cite{yamazaki:67}.  The overall anisotropy coefficients can then be deduced using Eq.~(\ref{eq:wcorr9}), thus, $A_{2}(2220) = 0.71429$ and $A_{4}(2220) = -1.71429$.  These results are also in exact agreement with the tabulated $B_{k}(J) F_{k}(J_{i}L_{1}L_{2}J_{f})$ values of Yamazaki \cite{yamazaki:67}.  Substituting these $A_{2}$ and $A_{4}$ coefficients into Eq.~(\ref{eq:wcorr3}) provides the theoretical angular-distribution correction factor as a function of angle assuming complete alignment in $^{56}$Fe for the $2^{+}_{1} \rightarrow 0^{+}_{\rm gs}$ transition.  This angular distribution function is presented in Fig.~\ref{fig:W_calc}.

\subsection{\label{sect:ang-s2} Experimental angular-distribution correction for the $2^{+}_{1} \rightarrow 0^{+}_{\rm gs}$ $\gamma$ ray in $^{56}$Fe\protect\\}

\begin{figure*}[t]
  \includegraphics[angle=0, width=\textwidth]{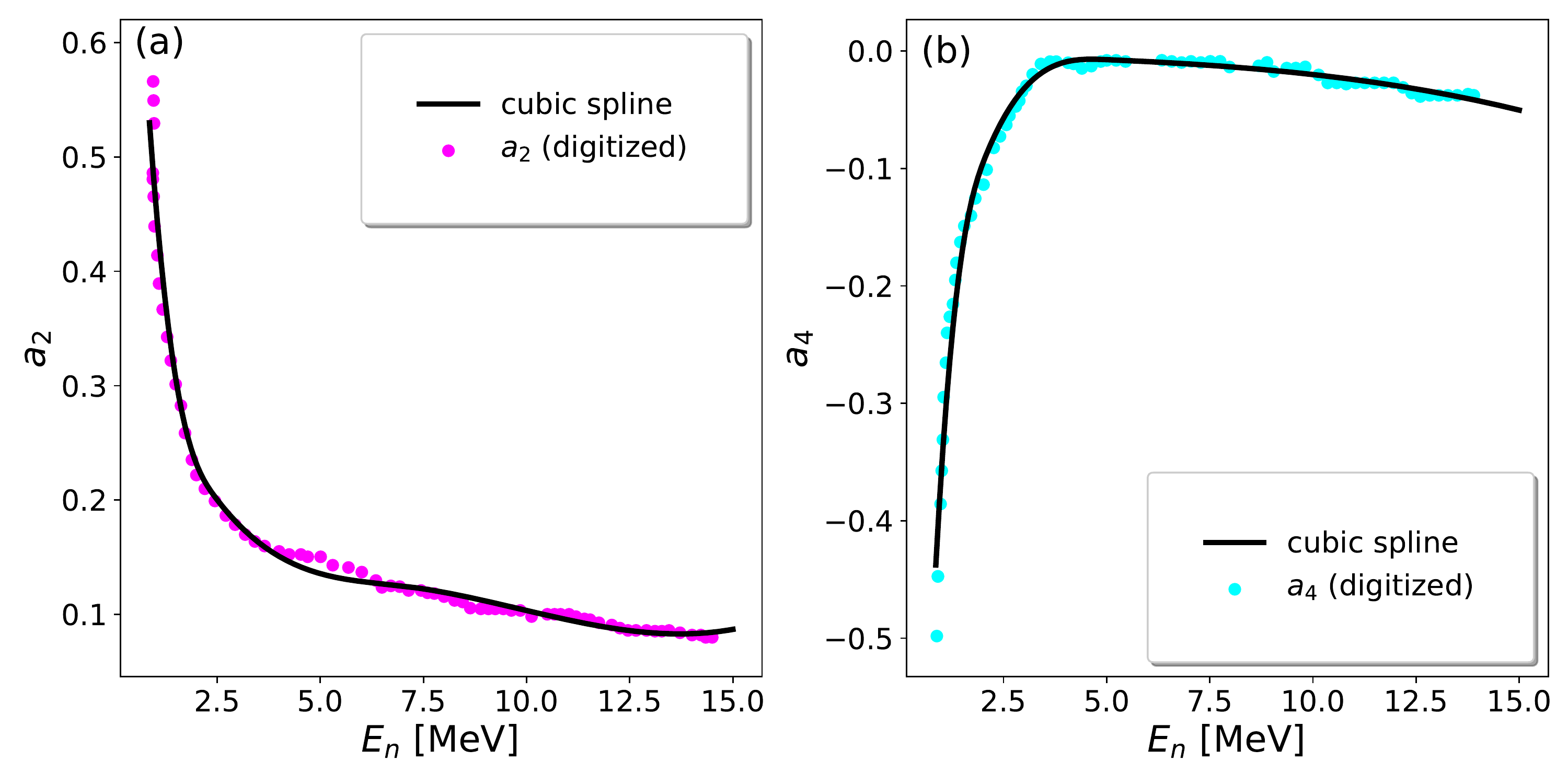}
  \caption{\label{fig:a2a4} Anisotropy-attenuation coefficients: (a) $a_{2}$, and (b) $a_{4}$, used to describe the angular distribution as a function of incident-neutron energy for the 846.8-keV $\gamma$ ray corresponding to the $2^{+}_{1} \rightarrow 0^{+}_{\rm gs}$ transition in $^{56}$Fe.  In each plot, the data points have been digitized from the fitted curves published by Savin \textit{et al}. \cite{savin:00}.  A cubic spline is used to provide a smooth interpolation throughout the range of the digitized sample.}
\end{figure*}

The theoretical scenario described in Sect.~\ref{sect:ang-s1} assumes complete nuclear alignment.  In real physical problems, the alignment is usually only partial and an anisotropy-attenuation coefficient $a_{k}$ is needed to describe the degree of alignment, where
\begin{equation}
  a_{k} = \frac{\rho_{k}(J)}{B_{k}(J)}.
  \label{eq:wexpt1}
\end{equation}
Longer-lived states tend to undergo greater amounts of deorientation leading to a greater attenuation in the alignment of the magnetic substates that will affect the measured $a_{k}$ values.  A correction factor due to the solid-angle subtended by the detector $Q_{k}$ also impacts the observed angular distribution such that Eq.~\ref{eq:wcorr3} may be recast using the more realistic expression
\begin{equation}
  W(\theta) = 1 + \sum\limits_{k=2,4} a_{k}Q_{k}P_{k}(\cos\theta).
  \label{eq:wexpt2}
\end{equation}
For a $4\pi$-array detector system $Q_{k} \rightarrow 0$ and the angular-distribution effect washes out, while at the other extreme, in the case of a point-like detector $Q_{k} \rightarrow 1$.  In this work, the single Ge(Li) detector positioned perpendicular to the beam line approximates that of a point-like detection system and thus, implies the correction factor $Q_{k} \approx 1$ in Eq.~(\ref{eq:wexpt2}).

Experimental anisotropy-attenuation coefficients are needed as a function of incident-neutron energy to describe the angular distribution for $^{56}$Fe($n,n'\gamma$).  To obtain these quantities, we have digitized the fitted $a_{2}$ and $a_{4}$ data for the 846.8-keV $\gamma$-ray transition from Ref.~\cite{savin:00} as shown in Fig.~\ref{fig:a2a4}.  These fitted experimental data were produced from a collation of several evaluated neutron-data libraries and experimental measurements over a range of incident-neutron energies from the threshold of the level-excitation energy up to 14~MeV.  The energy dependence reflects the fact that the population parameters of a state are highly dependent on the formation process of the state \cite{yamazaki:67}.  The effect of the incident-neutron energy on the expected angular distribution can be assessed by interpolating the $a_{k}$ coefficients in Fig.~\ref{fig:a2a4} using, e.g., a cubic spline interpolation method as shown, and calculating $W(\theta)$ in accordance with Eq.~(\ref{eq:wexpt2}).  Figure~\ref{fig:W_calc} illustrates experimental $W(\theta)$ functions using interpolated $a_{k}$ coefficients corresponding to neutron energies from the level-excitation threshold in $^{56}$Fe for the $2^{+}_{1}$ state up to 10.0~MeV.  The increasingly ``washed-out'' effect is apparent as the incident-neutron energy increases leading to increasingly oriented states that rapidly diverge from the theoretical case of maximum (or complete) nuclear alignment.

\begin{figure}[t]
  \includegraphics[width=\linewidth, angle=0]{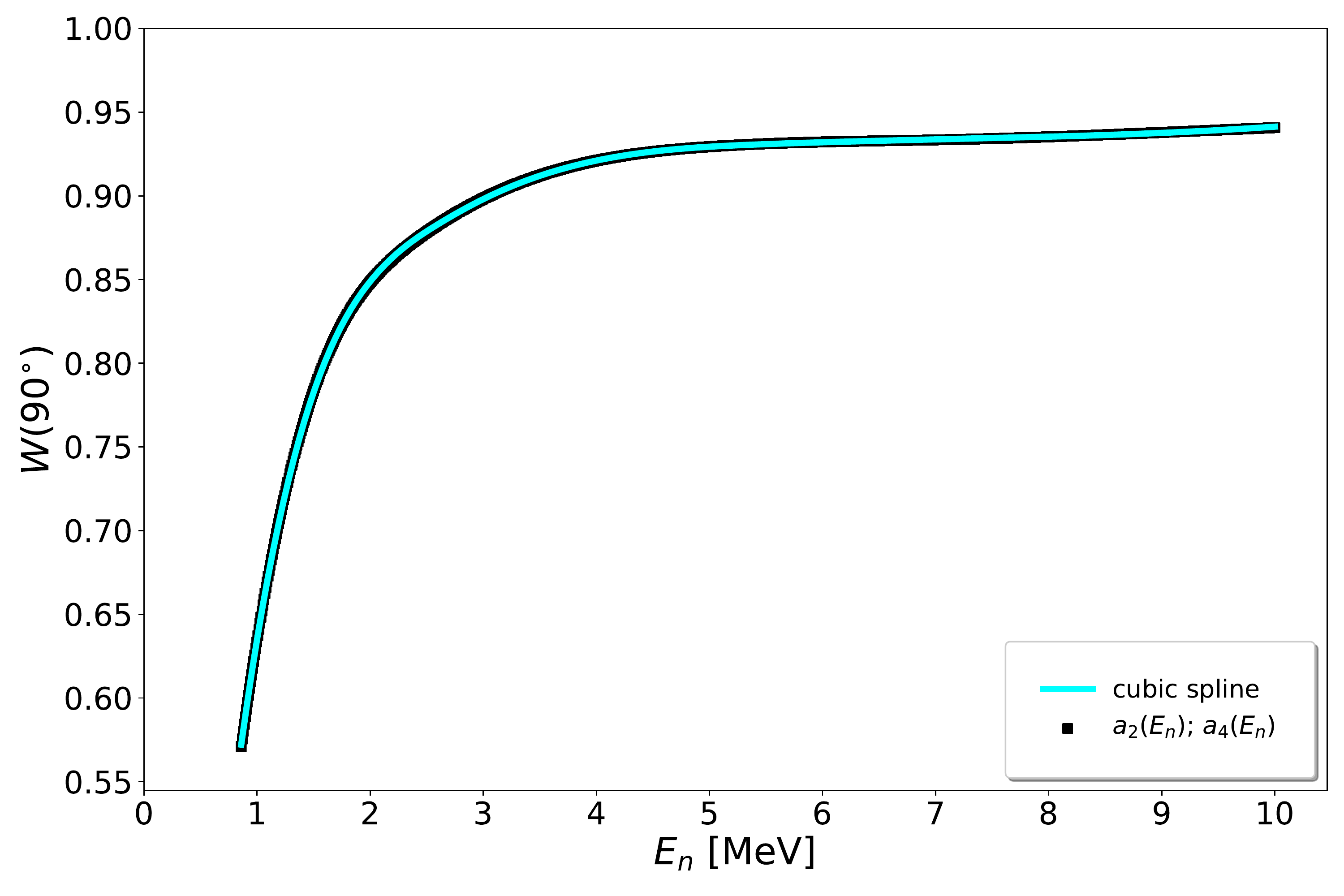}
  \caption{\label{fig:W_expt} Experimental $W(90^{\circ})$ function determined according to the interpolated $a_{k}$ coefficients \cite{savin:00} over a range of incident-neutron energies from level-excitation threshold to 14~MeV for the 846.8-keV $\gamma$-ray transition in $^{56}$Fe.  The fitted cubic spline used for interpolation in the analysis is shown.}
\end{figure}

The absolute $\gamma$-ray intensity data presented in the Atlas \cite{id:ATLAS} correspond to measurements at $90^{\circ}$ to the neutron beam line.  The extracted experimental $W(90^{\circ})$ correction factors, plotted in Fig.~\ref{fig:W_expt} as a function of incident-neutron energy, are therefore required to determine an appropriate angular-distribution corrected flux-weighted cross section for the 846.8-keV transition, as described in the next section.

\subsection{\label{sect:ang-s3} Angular distributions for other $\gamma$ rays\protect\\}

\begin{figure*}[ht]
  \includegraphics[angle=0, width=\textwidth]{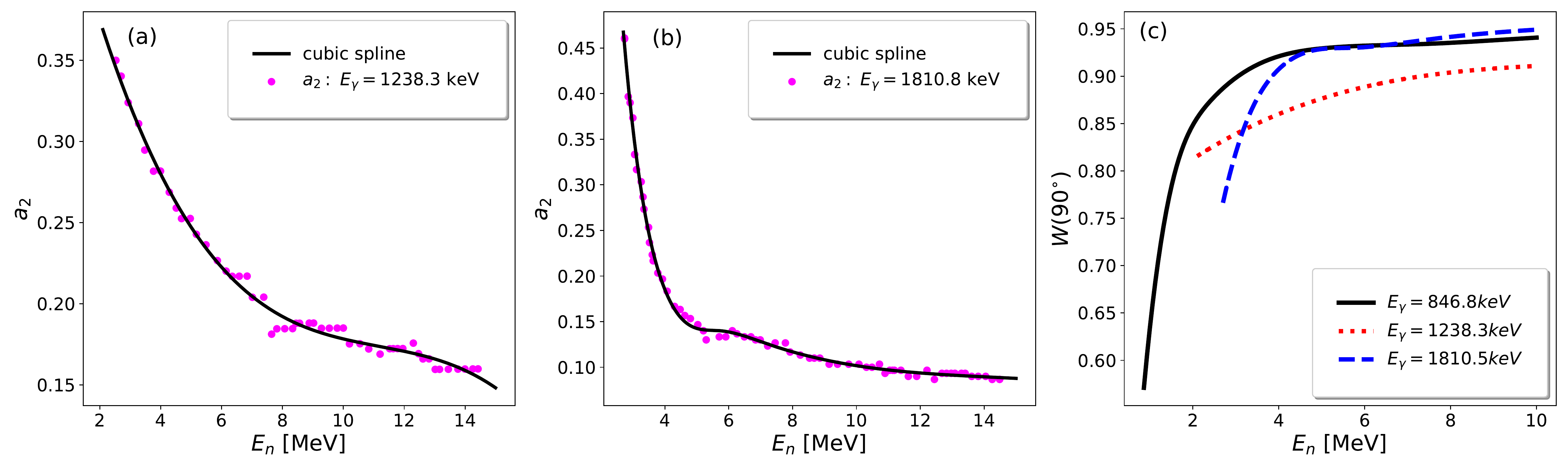}
  \caption{\label{fig:W_others} Experimental $a_{2}$ anisotropy coefficients \cite{savin:00} for $\gamma$-ray transitions in $^{56}$Fe at (a) 1238.3~keV, and (b) 1810.8~keV.  The $W(90^{\circ})$ functions, deduced according to interpolated experimental $a_{k}$ coefficients, are compared in (c) for the three strongest $\gamma$-ray transitions in $^{56}$Fe at 846.8, 1238.3, and 1810.8~keV.}
\end{figure*}

As shown in Sect.~\ref{sect:ang-s1}, the quantum mechanical properties associated with a $\gamma$-ray transition demonstrate that different $\gamma$ rays, even those described by a common radiation field, may have quite different angular distributions.  The experimental $a_{2}$ anisotropy coefficients for the 1238.3-keV ($4_{1}^{+} \rightarrow 2_{1}^{+}$; $E2$) and 1810.8-keV ($2_{2}^{+} \rightarrow 2^{+}_{1}$; $M1+E2$ with mixing ratio $\delta_{\gamma} = 0.18$ \cite{huo:11}) are shown in Figs.~\ref{fig:W_others}a and b, respectively, as a function of incident neutron energy.  Using the interpolated coefficients from Fig.~\ref{fig:W_expt} and Figs.~\ref{fig:W_others}a and b, the corresponding the $W(90^{\circ})$ functions for each of the three strongest transitions in $^{56}$Fe are shown in Fig.~\ref{fig:W_others}c.  Figure~\ref{fig:W_others}c shows that all three $\gamma$ rays have different angular distributions and, thus, mandate different correction factors as a function of neutron energy.


Because we are able to deduce an angular-distribution-corrected cross section for the 846.8-keV normalization transition ($\sigma_{\gamma_{N}}$), as described in the next section, this result can be used to generate scaled intensities ($\sigma_{\gamma_{i}}$) for all other $\gamma$-ray transition intensities ($I_{\gamma_{i}}$) reported in the Atlas according to,
\begin{equation}
  \label{eq:scale1}
  \sigma_{\gamma_{i}} = \sigma_{\gamma_{N}}I_{\gamma_{N}}I_{\gamma_{i}},
\end{equation}
where $I_{\gamma_{N}}$ is the $^{56}$Fe-relative-intensity normalization scaling factor, required in the two-stage normalization, for a given isotopic or elemental data set listed in Table~2 of Ref.~\cite{id:ATLAS}.  In reality, however, the angular distribution factor may be significant; preempting the discussion in the next section, we find corrections of the order of approximately $13-18\%$ are determined for the $^{56}$Fe $\gamma$-ray cross sections.  In fact, the authors of the Baghdad Atlas state that the correction may be as high as 30\% in some cases \cite{id:ATLAS}.  Thus, in general angular distribution corrections should be considered for the data presented in the Atlas and, therefore, Eq.~(\ref{eq:scale1}) should be modified to account for this correction
\begin{equation}
  \label{eq:scale2}
  \sigma_{\gamma_{i}} = \sigma_{\gamma_{N}}I_{\gamma_{N}}I_{\gamma_{i}}W_{\gamma_{i}}(90^{\circ}),
\end{equation}
where $W_{\gamma_{i}}(90^{\circ})$ is the corresponding angular-distribution correction factor for the transition $\gamma_{i}$.

\section{\label{sect:3} Cross-section calculations\protect\\}

The $\gamma$-ray intensities reported in the Baghdad Atlas \cite{id:ATLAS} have all been deduced using the two-stage normalization procedure described in Sect.~\ref{sect:2.2}. This method allows for all $I_{\gamma}(90^{\circ})$ values to be presented on an absolute scale relative to an assumed absolute intensity, or cross section, for the 846.8-keV $2^{+}_{1} \rightarrow 0^{+}_{\rm gs}$ transition in the $^{56}$Fe($n, n'\gamma$) reaction, as formulated by Eqs.~(\ref{eq:scale1}) and (\ref{eq:scale2}).  We have therefore deduced a value for the flux-weighted cross section for this transition, $\langle \sigma_{\gamma 847} \rangle$, to permit for the determination of partial $\gamma$-ray production cross sections for all Atlas-reported $I_{\gamma}(90^{\circ})$ data \cite{id:ATLAS}.  This expectation value can be determined by convolving the neutron flux spectrum shown in Figs.~\ref{fig:1} and \ref{fig:compound_func} with $\gamma$-ray production cross-section data corresponding to ($n, n'\gamma$), corrected for the angular distribution measured at $\theta = 90^{\circ}$ with respect to the neutron beam line.

\subsection{\label{sect:3.1} $\gamma$-ray production cross section\protect\\}

\begin{figure*}[t]
  \includegraphics[angle=0, width=\linewidth]{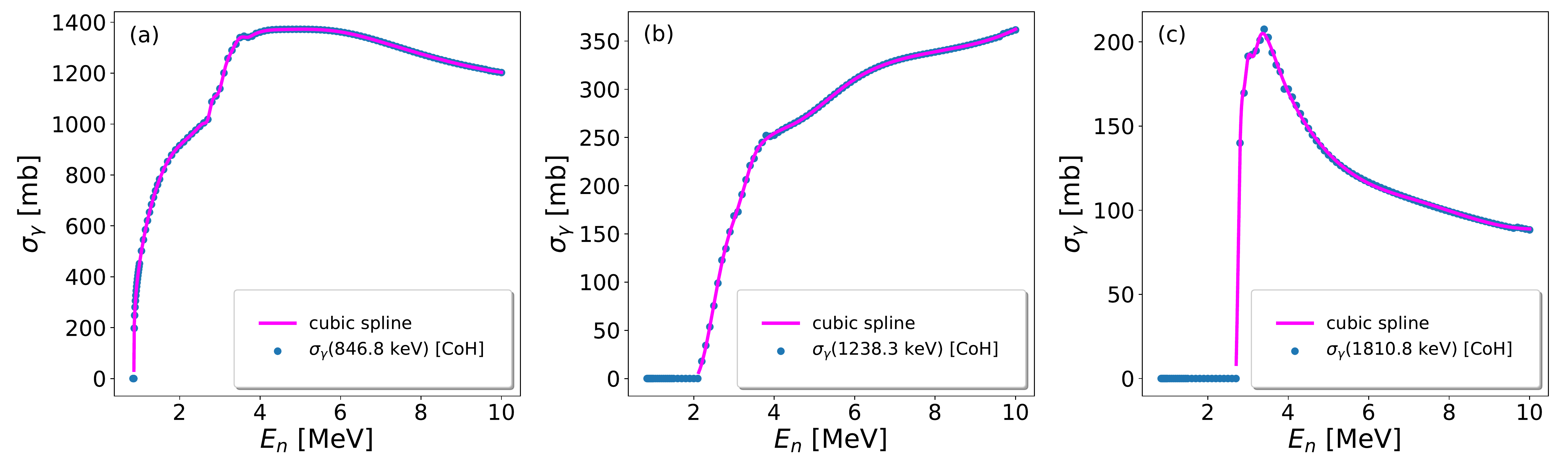}
  \caption{\label{fig:CoH} Calculated $\gamma$-ray production cross section as a function of the incident neutron energy obtained using the reaction code CoH$_{3}$ \cite{kawano:CoH} for $\gamma$-ray transitions in $^{56}$Fe at (a) 846.8~keV, (b) 1238.3~keV, and (c) 1810.8~keV.  See text for calculation details.}
\end{figure*}

The cross-section data available in neutron-data libraries, e.g., the Evaluated Nuclear Data File, ENDF/B-VIII.0 \cite{brown:18}, are deduced from models guided by experimental information where available.  One of the well-benchmarked reaction codes frequently used to provide modeled cross-section data is known as the Compact Optical model and Hauser-Feshbach code ({\small CoH$_{3}$}) \cite{kawano:CoH}.  The success of {\small CoH$_{3}$} in reliably predicting cross sections from inelastic-scattering reactions is well documented, e.g., Ref.~\cite{kerveno:19}.  This code is an optical model, exciton-preequilibrium and Hauser-Feshbach statistical-decay code that solves the Schr{\"o}dinger Equation for a defined set of optical model potentials and calculates the differential elastic-scattering, reaction, and total cross section for several incident light projectiles: $n, p, d, t$, $^{3}$He, and $\alpha$ particles.  In this work we have used {\small CoH$_{3}$} to calculate the production cross section for the 846.8-keV $\gamma$ ray over a range of incident-neutron energies from 0.87 to 10.00~MeV for the $^{56}$Fe($n,n'\gamma$) reaction.  These calculated $\gamma$-production cross sections are shown in Fig.~\ref{fig:CoH}.

To construct this calculation we have used optical model potentials describing the neutron \cite{koning:03}, proton \cite{koning:03}, and alpha \cite{avrigeanu:09} particles.  For the coupled-channels calculation, we have coupled all levels in $^{56}$Fe up to the 2-phonon triplet: 0.0 ($0^{+}_{1}$ ground state), 0.8468 ($2^{+}_{1}$ 1-phonon), 2.0851 ($4^{+}_{1}$ 2-phonon), 2.6576 ($2^{+}_{2}$ 2-phonon), and 2.9415~MeV ($0^{+}_{2}$ 2-phonon).  Additionally, we have assumed a deformation parameter $\beta = 0.23$ based on $(p,p')$ scattering measurements, where it is suggested that the first $2^{+}$ state in $^{56}$Fe is known to exhibit rotational behaviour \cite{delaroche:82}.  However, the calculations were found to be largely insensitive by adjusting the $\beta$ parameter to values expected for a near-spherical nucleus \cite{moller:95}.  The transmission coefficients determined from the optical model are then fed into the multistage statistical-model Hauser-Feshbach calculation.  For this component of the calculation, we have adopted: (i) the giant dipole resonance (GDR) parametrization of nearest-neighbouring even-even isotope $^{54}$Fe \cite{kawano:20} using experimental parametrizations to describe electric dipole $E1$ radiation: centroid $E_{G} = 19.39$~MeV; width $\Gamma_{G} = 5.69$~MeV; peak cross section $\sigma_{G} = 145.35$~mb, (ii) the GDR parametrization in accordance with the spin-flip model \cite{B&M:II} for magnetic dipole $M1$ radiation: $E_{G} = 41/A^{1/3}= 10.72$~MeV; $\Gamma_{G} = 4.0$~MeV; $\sigma_{G} = 1.0$~mb, and (iii) a global parametrization \cite{speth:81,prestwich:84} for the giant quadrupole resonance according to the isoscalar plus isovector model of the photon strength function to describe the electric quadrupole $E2$: $E_{G} = 63/A^{1/3} = 16.47$~MeV; $\Gamma_{G} = 6.11 - 0.012A = 5.44$~MeV; $\sigma_{G} = 1.5 \times 10^{-4} \frac{Z^{2} E_{G}^{2}}{A^{1/3}\Gamma_{G}} = 1.32$~mb.  The {\small CoH$_{3}$} calculation method also includes a two-component exciton model to describe pre-equilibrium particle emissions.  Default two-body matrix elements controlling the pre-equilibrium emission strength are used \cite{kawano:CoH}.  Finally, the {\small CoH$_{3}$} calculations also invoke a direct/semi-direct capture model, whereupon the default parametrizations used to describe the non-locality correction for the distorted wave (empirically known to be 0.85) together with real and imaginary potentials describing the coupling strength of the semi-direct process, have been adopted \cite{kawano:CoH}.

\subsection{\label{sect:3.2} Flux-weighted cross section\protect\\}

\begin{figure*}[ht]
  \includegraphics[angle=0, width=\linewidth]{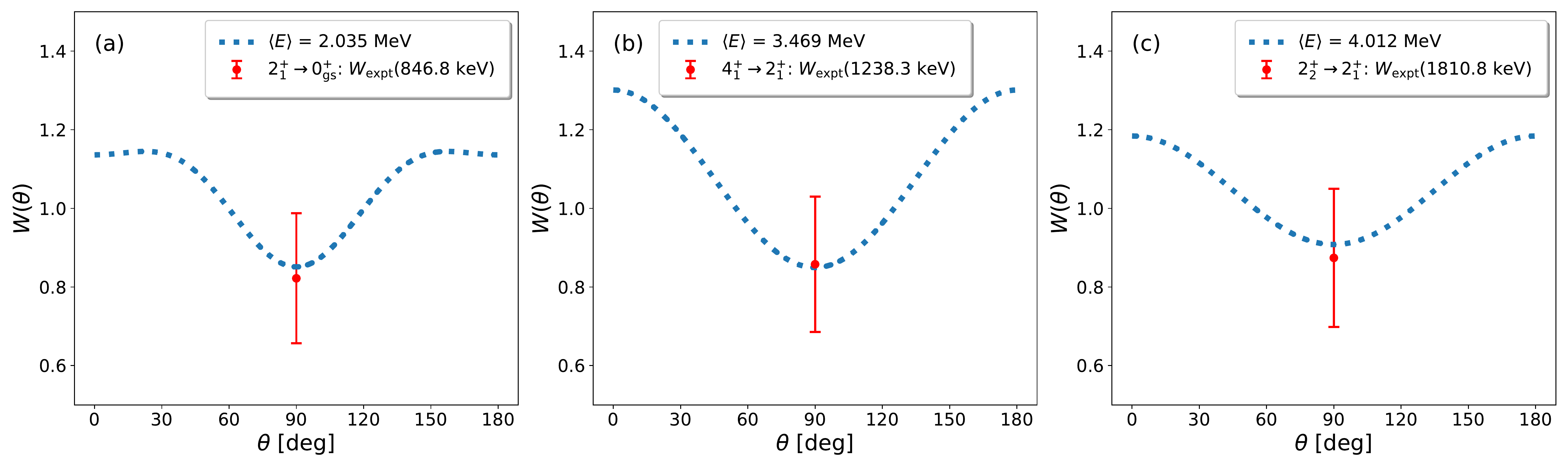}
  \caption{\label{fig:W_expt_dists} Experimentally-deduced $W(90^{\circ})$ correction factors (data points) compared to the overall $W(\theta)$ angular distribution function (dotted curves) corresponding to the deduced flux-weighted neutron-expectation energy giving rise to the three strongest $^{56}$Fe $\gamma$ rays at (a) 846.8~keV [$2^{+}_{1} \rightarrow 0^{+}_{\rm gs}$; $E2$], (b) 1238.3~keV [$4^{+}_{1} \rightarrow 2^{+}_{1}$; $E2$], and (c) 1810.8~keV [$2^{+}_{2} \rightarrow 2^{+}_{1}$; $M1+E2$].}
\end{figure*}

The flux-weighted cross section for the 846.8-keV $2^{+}_{1} \rightarrow 0^{+}_{\rm gs}$ $\gamma$-ray transition corrected for the angular distribution can be determined according to
\begin{table}[h]
  \caption{\label{tab:integrals} Definite and numerical-approximation flux integrals corresponding to the measured spectrum reported in the Baghdad Atlas \cite{id:ATLAS, ahmed:74A} according to the parametrization of Eq.~(\ref{eq:phi_overall}) deduced in this work.  The method column refers to the adopted {\tt scipy} \cite{id:SciPy} integration method.  The results shown in the final column are in arbitrary units.}
  \begin{tabular}{ccc} \hline\hline
    Method & Type & $\int\limits_{0}^{+\infty} \phi(E)dE$ \\ \hline
    {\tt quad} & Analytical; definite & 88,737\\
    {\tt trapz} & Trapezium ; numerical & 88,658\\
    {\tt simps} & Simpson; numerical & 88,689\\
    \hline\hline
  \end{tabular}
\end{table}
\begin{equation}
  \langle \sigma_{\gamma} \rangle = \frac{\int\limits^{10.00}_{E_{\rm th}} \phi(E) \sigma_{\gamma}(E) W_{\gamma}(\theta=90^{\circ};E) dE}{\int\limits^{+\infty}_{0} \phi(E) dE},
  \label{eq:wexpt3}
\end{equation}
where the lower limit $E_{\rm th}$ is the reaction threshold for an energy level with an excitation energy $E_{L}$ and is deduced as
\begin{equation}
  E_{\rm th} = E_{L} \frac{A+1}{A},
  \label{eq:e_thresh}
\end{equation}
and thus, $E_{\rm th} = 0.862$~MeV for the $2^{+}_{1}$ state.  We have determined values for the denominator in Eq.~(\ref{eq:wexpt3}) using numerical integration methods: both the trapezium method and the Simpson method yield statistically consistent results.  As an additional check, we have also calculated this result analytically using the {\tt quad} method from the {\tt scipy.integrate} \cite{id:SciPy} third-party {\tt Python} \cite{id:Python} library, which again gives a consistent result to those obtained using the aforementioned approximation methods.  The results of these integration methods are tabulated and compared in Table~\ref{tab:integrals}.

Convolution of the flux spectrum $\phi(E)$ given by Eq.~(\ref{eq:phi_overall}) with the {\small CoH$_{3}$}-calculated cross section in Fig.~\ref{fig:CoH}a together with the angular-distribution correction function in Fig.~\ref{fig:W_expt} permits for the evaluation of the numerator in Eq.~(\ref{eq:wexpt3}), above.  Using this method we find $\langle \sigma_{\gamma} \rangle = 200 (41)$~mb.  The overall uncertainty is dominated by an estimated 20\% contribution from the {\small CoH$_{3}$} calculations \cite{kawano:PC}, combined with an additional 2\% contribution \cite{savin:00} assumed for the angular-distribution correction.  In the absence of the angular distribution correction, however, Eq.~(\ref{eq:wexpt3}) reduces to
\begin{equation}
  \langle \sigma_{\gamma} \rangle = \frac{\int\limits^{10.00}_{E_{\rm th}} \phi(E) \sigma_{\gamma}(E) dE}{\int\limits^{+\infty}_{0} \phi(E) dE},
  \label{eq:wexpt4}
\end{equation}
and the uncorrected cross section becomes $\langle \sigma_{\gamma} \rangle = 244(49)$~mb.  As shown by Eq.~(\ref{eq:wcorr2}), the ratio of the corrected to uncorrected flux-weighted cross sections provides an average angular distribution correction factor, in this case $W(90^{\circ}) = 0.82(16)$.  Here, we have assumed an overall uncertainty of only 20.1\% due to the highly-correlated uncertainties in the {\small CoH$_{3}$} calculations.  This result is presented in Fig.~\ref{fig:W_expt_dists}a.

\begin{table*}[h]
  \caption{\label{tab:fwcs} Flux-weighted $\gamma$-ray production cross sections, with [$\langle \sigma_{\gamma} \rangle_{W}$; Eq.~(\ref{eq:wexpt3})] and without [$\langle \sigma_{\gamma} \rangle$; Eq.~(\ref{eq:wexpt4})] correction to the $\gamma$-ray angular distribution.  The thresholds [$E_{\rm th}$, Eq.~(\ref{eq:e_thresh})] and corresponding initial excitation energies ($E_{L}$) are listed for each $\gamma$ ray ($E_{\gamma}$).  The Atlas branching ratios ($B_{A}$) are deduced using the $I_{\gamma}$ \cite{id:ATLAS} data and the $B_{kT}$ branching ratios have been deduced assuming $kT=0.39$~MeV from the parametrization of the Atlas-reported neutron flux $\phi(E)$ given by Eq.~(\ref{eq:phi_overall}).  Residuals ($R$) are listed in the final column.}
  \begin{center}
    \begin{tabular}{ccccccccc} \hline\hline
      $E_{\gamma}$ [keV] & $E_{L}$ [keV] & $E_{\rm th}$ [keV] & $I_{\gamma}$ [\%] & $B_{A}$ & $\langle \sigma_{\gamma} \rangle_{W}$ [mb] & $\langle \sigma_{\gamma} \rangle$ [mb] & $B_{kT}$ & $R$ [$\sigma$] \\
      \hline
      846.8 & 846.8 & 861.9 & 100.0 & 1.0 & 200(41) & 244(49) & 1.0 & $-$\\
      1238.3 & 2085.1 & 2122.3 & 10.5(5) & 0.105(5) & 14.9(30) & 17.4(35) & 0.075(21) & 1.4 \\
      1810.8 & 2657.6 & 2705.1 & 6.9(4) & 0.069(4) & 9.5(19) & 10.9(22) & 0.048(14) & 1.4 \\

      \hline\hline
    \end{tabular}
  \end{center}
\end{table*}

Similarly, we have used Eqs.~(\ref{eq:wexpt3}) to (\ref{eq:wexpt4}) to deduce flux-weighted cross sections for the other two strongest $\gamma$ rays in $^{56}$Fe, with and without the correction for the angular distribution (see Fig.~\ref{fig:W_others}c for the experimental $W(90^{\circ})$ angular distribution functions), at 1238.3~keV ($4^{+}_{1} \rightarrow 2_{1}^{+}$; $E2$) and 1810.8~keV ($2^{+}_{2} \rightarrow 2_{1}^{+}$; $M1+E2$).  The {\small CoH$_{3}$}-calculated cross sections used for these two $\gamma$ rays are shown in Figs.~\ref{fig:CoH}b and c, respectively, and the corresponding flux-weighted cross sections are listed, together with the 846.8-keV result, in Table~\ref{tab:fwcs}.  Our flux-weighted cross sections are compared to the original $I_{\gamma}$ data from the Atlas \cite{id:ATLAS} by calculating branching ratios for the 1238.3- and 1810.8-keV $\gamma$ rays relative to the 846.8 keV.  The Atlas branching ratios are denoted $B_{A}$ while our values, deduced using the flux parametrization of $kT=0.39$ MeV and $\beta=0.76$ to describe the measured neutron flux $\phi(E)$ given by Eq.~(\ref{eq:phi_overall}) in Sect.~\ref{sect:2.1}, are denoted $B_{kT}$.  Explicitly, these branching ratios are calculated according to
\begin{equation}
  \label{eq:br}
  B_{A_{i}} = \frac{I_{\gamma_{i}}}{I_{\gamma 846.8}} \qquad {\rm and} \qquad B_{kT_{i}} = \frac{\langle \sigma_{\gamma_{i}} \rangle_{W}}{\langle \sigma_{\gamma_{846.8}} \rangle_{W}}.
\end{equation}
We can then assess the differences between these results by calculating the residuals in units of standard deviation [$\sigma$] assuming
\begin{equation}
  \label{eq:res}
  R = \frac{|B_{A} - B_{kT}|}{\sqrt{dB_{A}^{2} + dB_{kT}^{2}}}.
\end{equation}
Here, we find our flux-weighted measurements agree with the Atlas-reported branching ratios at the 1.4$\sigma$ level, implying reasonable consistency.  Furthermore the intensity ratio for the high-energy $\gamma$ rays, $I_{\gamma 1283.3}/I_{\gamma 1810.8} = 1.52(11)$, is in excellent agreement with the corresponding cross-section ratio, $\langle \sigma_{\gamma 1283.3} \rangle / \langle \sigma_{\gamma 1810.8} \rangle = 1.57(45)$.  Because a neutron-energy threshold of $E > 2.1$~MeV is required to induce observation of these $\gamma$ rays, this result suggests that our parameterized neutron-flux spectrum for $\phi(E)$ is well understood in the region $E \gtrsim 2$~MeV.  However, at lower-neutron energies further optimization of the flux parametrization is required, as described in the forthcoming Sect.~\ref{sect:3.6}.


\subsection{\label{sect:3.3} Flux-weighted expectation energy\protect\\}

Using a similar procedure to that outlined above, and considering only neutron energies that lead to excitation of a particular $\gamma$-ray-emitting state in $^{56}$Fe listed in Table~\ref{tab:fwcs}, we can deduce the corresponding expectation value of the incident-neutron energy spectrum with the following relationship
\begin{equation}
  \langle E \rangle = \frac{\int\limits^{10.00}_{E_{\rm th}} \phi(E) E dE}{\int\limits^{10.00}_{E_{\rm th}} \phi(E) dE}.
  \label{eq:wexpt5}
\end{equation}
Again, using both the trapezium and Simpson methods to evaluate the integrals in Eq.~(\ref{eq:wexpt5}), we find statistically consistent results for the deduced expectation energies $\langle E \rangle$.  These results are presented in Table~\ref{tab:energies}.  Interpolation of the $a_{k}$ attenuation-anisotropy coefficients (see Figs.~\ref{fig:a2a4} and \ref{fig:W_others}) at the deduced $\langle E \rangle$ provides the appropriate the angular-distribution function, whereupon the corresponding correction factor $W(90^{\circ})$ can also be interpolated.  These $\langle E \rangle$-deduced angular-distribution functions are shown in Fig.~\ref{fig:W_expt_dists} in comparison to the experimentally-deduced $W(90^{\circ})$ values determined in accordance with Eq.~(\ref{eq:wcorr2}) for each of the three $\gamma$ rays listed in Table~\ref{tab:energies}.  The plots reveal consistency between both approaches.  The $\langle E \rangle$-interpolated $W(90^{\circ})$ are also presented in Table~\ref{tab:energies} in comparison to those deduced using Eq.~(\ref{eq:wcorr2}).  These results generally indicate that as the neutron energy increases, the angular-distribution correction factor also increases, i.e., $\sigma(\theta) \rightarrow \sigma_{\rm inl}$ cf. Eq.~(\ref{eq:wcorr2}).

\begin{table}[hb]
  \caption{\label{tab:energies} Neutron expectation energies ($\langle E \rangle$) above the threshold ($E_{\rm th}$) giving rise to the listed $\gamma$ rays ($E_{\gamma}$).  The $W(90^{\circ})_{I}$ correction factors are interpolated from Fig.~\ref{fig:W_expt_dists} and the $W(90^{\circ})_{E}$ correction factors are deduced using Eq.~(\ref{eq:wcorr2}) together with the extracted cross sections listed in Table~\ref{tab:fwcs}. }
  \begin{center}
    \begin{tabular}{ @{\hspace{0.75\tabcolsep}} c @{\hspace{0.75\tabcolsep}} c @{\hspace{0.75\tabcolsep}} c @{\hspace{0.75\tabcolsep}} c @{\hspace{0.75\tabcolsep}} c} \hline\hline
      $E_{\gamma}$ [keV] & $E_{\rm th}$ [keV] & $\langle E \rangle$ [MeV] & $W(90^{\circ})_{I}$ & $W(90^{\circ})_{E}$ \\
      \hline
      846.8 & 861.9 & 2.035 & 0.852 & 0.82(17) \\
      1238.3 & 2122.3 & 3.469 & 0.850 & 0.86(17) \\
      1810.8 & 2705.1 & 4.012 & 0.909 & 0.87(18) \\
      \hline\hline
    \end{tabular}
  \end{center}
\end{table}

\begin{figure*}[ht]
  \includegraphics[angle=0, width=\linewidth]{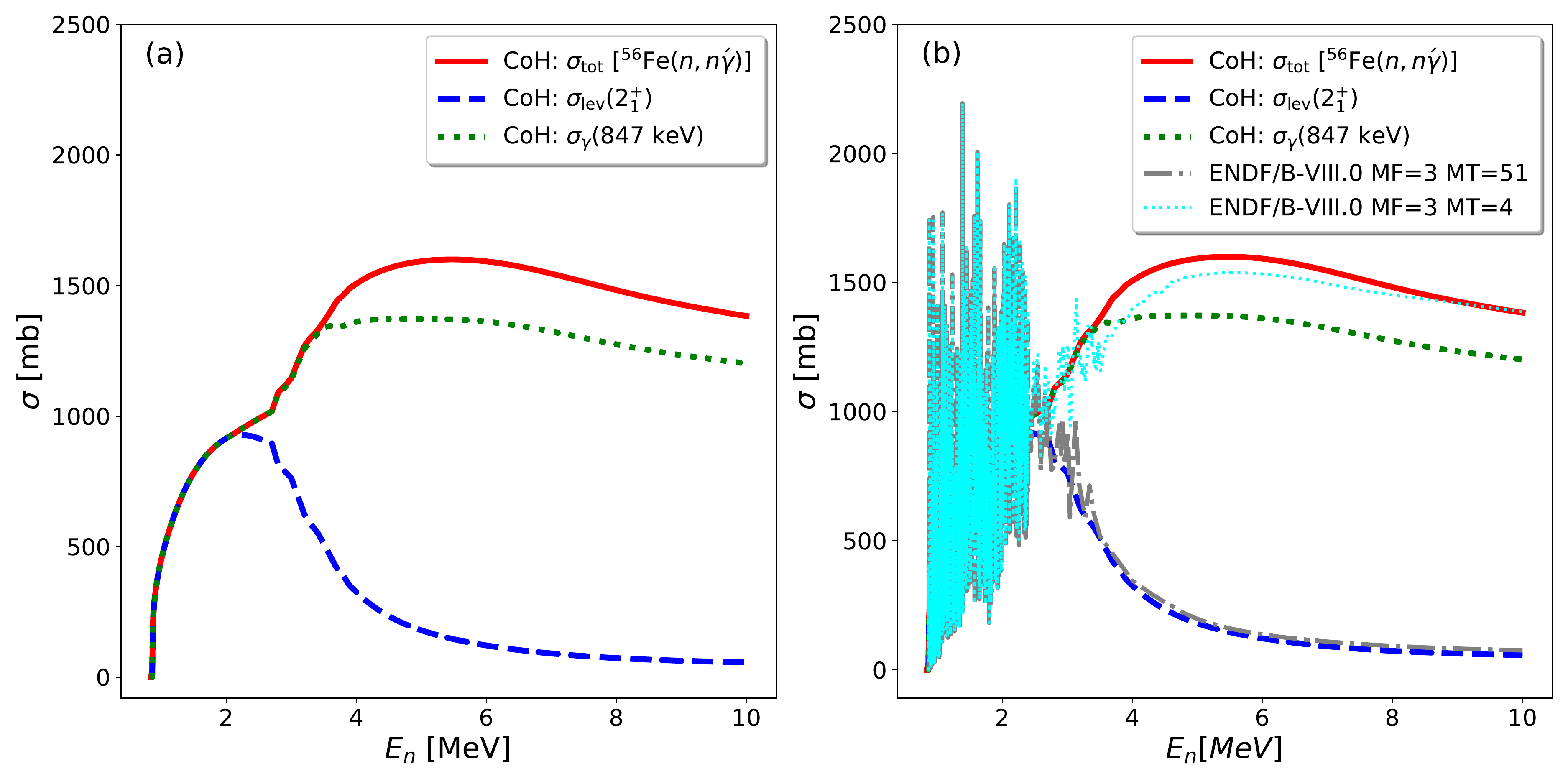}
  \caption{\label{fig:CoH-ENDF} (a) Calculated cross sections obtained using CoH$_{3}$ \cite{kawano:CoH}: total inelastic cross section for the $^{56}$Fe($n,n'\gamma$) reaction (solid red curve); direct excitation to the first $2^{+}$ state from inelastic scattering (dashed blue curve); $\gamma$-ray production cross section for the $2^{+}_{1} \rightarrow 0^{+}_{\rm gs}$ 846.8-keV transition (dotted green curve) in $^{56}$Fe.  (b) The CoH$_{3}$ calculations overlaid with the corresponding ENDF/B-VIII.0 \cite{brown:18}data: {\tt MF=3 MT=4} represents the total inelastic cross section (dotted cyan curve); {\tt MF=3 MT=51} represents the direct excitation function for the $2^{+}_{1}$ state in $^{56}$Fe (dashed-dotted grey curve).}
\end{figure*}

\subsection{\label{sect:3.4} Comparison with evaluated data\protect\\}

To help verify the integrity of the calculated cross sections in this paper, we have compared our calculations to the evaluated cross-section data from ENDF/B-VIII.0 \cite{brown:18}.  Figure~\ref{fig:CoH-ENDF}a shows the following quantities calculated using {\small CoH$_{3}$} \cite{kawano:CoH}: (i) the total inelastic scattering cross section for the $^{56}$Fe($n,n'\gamma$) reaction ($\sigma_{\rm tot}^{\rm CoH}$); (ii) the direct excitation function corresponding to the inelastic cross section for the first-excited state ($J^{\pi} = 2^{+}, E=846.8$~keV) in $^{56}$Fe ($\sigma_{\rm lev}^{\rm CoH}$); (iii) the $\gamma$-ray production cross section for the $2^{+}_{1} \rightarrow 0^{+}_{\rm gs}$ transition at 846.8~keV ($\sigma_{\gamma}^{\rm CoH}$).  In Fig.~\ref{fig:CoH-ENDF}b, these results are compared to the corresponding cross-section data from ENDF/B-VIII.0: the total inelastic-scattering cross section for the $^{56}$Fe($n,n'\gamma$) reaction in {\tt MF=3 MT=4} ($\sigma_{\rm tot}^{\rm ENDF}$); direct excitation function for the first $2^{+}$ state in {\tt MF=3 MT=51} ($\sigma_{\rm lev}^{\rm ENDF}$).  Unfortunately, ENDF does not provide partial $\gamma$-ray production cross sections directly making it difficult to compare to the calculated $\gamma$-production cross section determined with {\small CoH$_{3}$} for the 846.8-keV transition.  However, it is encouraging to see that ENDF data for both the total-reaction cross section and the direct-production cross section of the $2^{+}$ state track closely the cross sections calculated using {\small CoH$_{3}$}.  Accordingly, we may use this information to obtain an estimate of the $\gamma$-ray production cross section from ENDF ($\sigma_{\gamma}^{\rm ENDF}$).

The {\small CoH$_{3}$} results and the data from ENDF in Fig.~\ref{fig:CoH-ENDF} implies that
\begin{equation}
  \label{eq:cf1}
  \frac{\langle \sigma_{\rm lev}^{\rm CoH} \rangle}{\langle \sigma_{\rm tot}^{\rm CoH} \rangle} \approx \frac{\langle \sigma_{\rm lev}^{\rm ENDF} \rangle}{\langle \sigma_{\rm tot}^{\rm ENDF} \rangle}.
\end{equation}
The flux-weighted quantities in Eq.~(\ref{eq:cf1}) can be determined in a similar manner outlined earlier cf. Eq.~(\ref{eq:wexpt4}).  Adopting this procedure we find for the {\small CoH$_{3}$}-weighted cross sections $\langle \sigma_{\rm lev}^{\rm CoH} \rangle = 186.4$~mb and $\langle \sigma_{\rm tot}^{\rm CoH} \rangle = 250.3$~mb, thus, $\langle \sigma_{\rm lev}^{\rm CoH} \rangle / \langle \sigma_{\rm tot}^{\rm CoH} \rangle = 0.74$.  Similarly for the ENDF-weighted cross sections $\langle \sigma_{\rm lev}^{\rm ENDF} \rangle = 175.5$~mb and $\langle \sigma_{\rm tot}^{\rm ENDF} \rangle = 232.0$~mb, therefore, $\langle \sigma_{\rm lev}^{\rm ENDF} \rangle / \langle \sigma_{\rm tot}^{\rm ENDF} \rangle = 0.76$.  Because these ratios are in agreement in accordance with Eq.~(\ref{eq:cf1}), it is reasonable to expect the ratios of the $\gamma$-ray production cross section to the total reaction cross section should also be in agreement.  We can, therefore, estimate the $\gamma$-ray production from ENDF as
\begin{equation}
  \langle \sigma_{\gamma}^{\rm ENDF} \rangle \approx \int\limits_{0}^{+\infty} \phi(E) \sigma_{\rm tot}^{\rm ENDF}(E) \frac{\sigma_{\gamma}^{\rm CoH}(E)}{\sigma_{\rm tot}^{\rm CoH}(E)} dE,
  \label{eq:cf2}
\end{equation}
or, alternatively
\begin{equation}
  \langle \sigma_{\gamma}^{\rm ENDF} \rangle \approx \langle \sigma_{\rm tot}^{\rm ENDF} \rangle \frac{\langle \sigma_{\gamma}^{\rm CoH} \rangle}{\langle \sigma_{\rm tot}^{\rm CoH} \rangle}.
  \label{eq:cf3}
\end{equation}
Using the results for $\langle \sigma_{\gamma} \rangle$ determined in Sect.~\ref{sect:3.2}, together with flux-weighted values from above, Eq.~(\ref{eq:cf3}) gives $\langle \sigma_{\gamma}^{\rm ENDF} \rangle = 185.7$~mb corrected for the $\gamma$-ray angular distribution, and $\langle \sigma_{\gamma}^{\rm ENDF} \rangle = 225.9$~mb without the angular-distribution correction.  Both estimates are in close agreement with those obtained using {\small CoH$_{3}$}.

\subsection{\label{sect:3.5} Correction for $^{56}$Fe($n,p$)$^{56}$Mn\protect\\}

\begin{figure}[t]
  \includegraphics[angle=0, width=\linewidth]{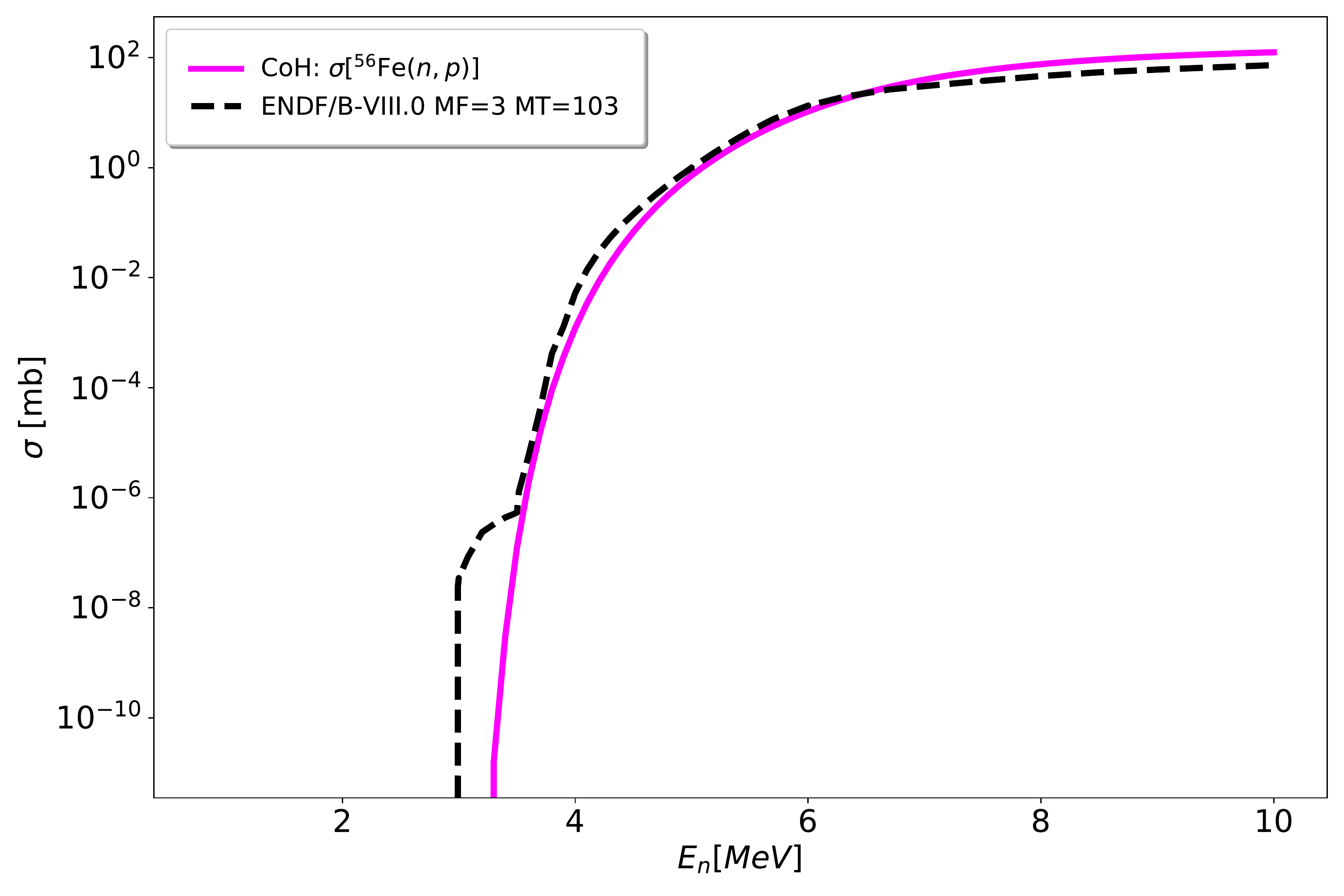}
  \caption{\label{fig:np} Cross sections for the $^{56}$Fe($n,p$) reaction: calculated using CoH$_{3}$ (solid magenta curve) and taken from the {\tt MF=3 MT=103} file from the ENDF/B-VIII.0 library (dashed black line).}
\end{figure}

There is an additional contribution to the 846.8-keV $\gamma$-ray yield from the $^{56}$Fe($n,p$) channel due to the $\beta^{-}$ decay of $^{56}$Mn to the $2^{+}_{1}$ state in $^{56}$Fe.  In the original work of Ref.~\cite{id:ATLAS}, the authors were unable to discriminate between ($n,n'$) and ($n,p$) channels and so the measured yield for the 846.8-keV $\gamma$ ray ($I_{\gamma_{\rm tot}}$) accounts for both contributions:
\begin{equation}
  \label{eq:Mn}
  I_{\gamma_{\rm tot}} = I_{\gamma_{nn}} + I_{\gamma_{np}},
\end{equation}
where $I_{\gamma_{nn}}$ is the $\gamma$ production for the 846.8-keV transition from $^{56}$Fe($n,n'\gamma$) and $I_{\gamma_{np}}$ is from $^{56}$Fe($n,p$).  It is important, therefore, to assess the significance of the ($n,p$) contribution.  Figure~\ref{fig:np} shows both the {\small CoH$_{3}$}-calculated cross sections and the ENDF/B-VIII.0 cross-section data from {\tt MF=3 MT=103} for the $^{56}$Fe($n,p$) reaction.  The calculated flux-weighted cross sections are determined to be $\langle \sigma_{np}^{\rm CoH} \rangle = 0.54$~mb and $\langle \sigma_{np}^{\rm ENDF} \rangle = 0.47$~mb, hence, in close agreement.  These results are statistically insignificant in comparison to the flux-weighted cross sections from the ($n,n'\gamma$) channel, $\langle \sigma_{np}^{\rm CoH} \rangle$ is less than 0.3\% that of $\langle \sigma_{\gamma} \rangle$, and no further correction to the $\gamma$-ray yield reported in Ref.~\cite{id:ATLAS} is required.

Throughout the Baghdad Atlas, for most $\gamma$-ray transitions in most nuclei $I_{\gamma_{\rm tot}} \approx I_{\gamma_{nn}}$ \cite{id:ATLAS}.  However, for the four to six lowest-lying levels \cite{id:ATLAS}, such corrections may need to considered for nuclei where additional reaction channels, other than ($n,n'\gamma$),  may pose a significant source of contamination.

\subsection{\label{sect:3.6} Neutron-flux optimization\\}

\begin{table*}[t]
  \caption{\label{tab:fwa_opt} Optimized flux-weighted angular-distribution-corrected $\gamma$-ray production cross sections ($\langle \sigma_{\gamma} \rangle_{W}$) for the three strongest transitions in $^{56}$Fe, deduced assuming the best-fit value $kT=0.155$~MeV.  For comparison, the corresponding average flux-weighted cross sections ($\langle \sigma_{\gamma} \rangle_{\rm FRM}$) obtained from the FRM-II measurement \cite{ilic:20} are also listed.  The final column shows the Baghdad Atlas cross sections scaled by the ratio of expectation energies deduced from the two facilities in accordance with Eq.~(\ref{eq:frm}).}
  \begin{center}
    \begin{tabular}{ @{\hspace{3.0\tabcolsep}} c @{\hspace{3.0\tabcolsep}} c @{\hspace{3.0\tabcolsep}} c @{\hspace{3.0\tabcolsep}} c @{\hspace{3.0\tabcolsep}} c @{\hspace{3.0\tabcolsep}} c @{\hspace{3.0\tabcolsep}} c} \hline\hline
      $E_{\gamma}$ [keV] & $B_{A}$ & $B_{kT}$ & $R$ [$\sigma$] & $\langle \sigma_{\gamma} \rangle_{W}$ [mb] & $\langle \sigma_{\gamma} \rangle_{{\rm FRM}}$ [mb] & $\langle \sigma_{\gamma} \rangle_{S}$ [mb]  \\
      \hline
      846.8  & 1.0       & 1.0       & $-$   & 143(29)  & 586(41) & 521(106) \\
      1238.3 & 0.105(5)  & 0.096(27) & 0.33  & 13.7(27) & 58(5)   & 49.9(98) \\
      1810.8 & 0.069(4)  & 0.061(18) & 0.44  & 8.7(18)  & 37(3)   & 31.7(66) \\
      \hline\hline
    \end{tabular}
  \end{center}
\end{table*}

The residuals in Table~\ref{tab:fwcs} already demonstrate reasonable agreement between the Atlas-reported intensities \cite{id:ATLAS} and the flux-weighted cross sections, assuming $kT=0.39$~MeV to describe the low-energy Maxwellian component of $\phi(E)$ in Eq.~(\ref{eq:phi_overall}).  However, we may also treat $kT$ as an adjustable parameter in the determination of the flux-weighted branching ratios $B_{kT}$, given by Eq.~(\ref{eq:br}), in order to optimize agreement with the corresponding Atlas branching ratios $B_{A}$.  Because the $\gamma$-ray intensities (or cross sections) are correlated, a covariance matrix $\bm{V}$ is needed to describe the $\chi^{2}$ minimization
\begin{equation}
  \label{eq:chi2_1}
  \chi^{2} = \sum\limits_{i=1}^{N}\sum\limits_{j=1}^{N} [B_{A_{i}}-B_{kT_{i}}][V_{ij}^{-1}][B_{A_{j}}-B_{kT_{j}}],
\end{equation}
where $N=3$ represents the number of $\gamma$ rays used in the minimization procedure.  Writing $\bm{B_{A}}$ and $\bm{B_{kT}}$ as vectors of $N$ elements each, we may then recast the $\chi^{2}$ distribution in matrix notation as
\begin{equation}
  \label{eq:chi2_2}
  \chi^{2} = (\bm{B_{A}}-\bm{B_{kT}}) \bm{V}^{-1} (\bm{\widetilde{B_{A}}}-\bm{\widetilde{B_{kT}}}),
\end{equation}
where $\bm{\widetilde{B_{A}}}$ and $\bm{\widetilde{B_{kT}}}$ are the transposed vectors of $\bm{B_{A}}$ and $\bm{B_{kT}}$, respectively.

\begin{figure*}[t]
  \includegraphics[width=\linewidth, angle=0]{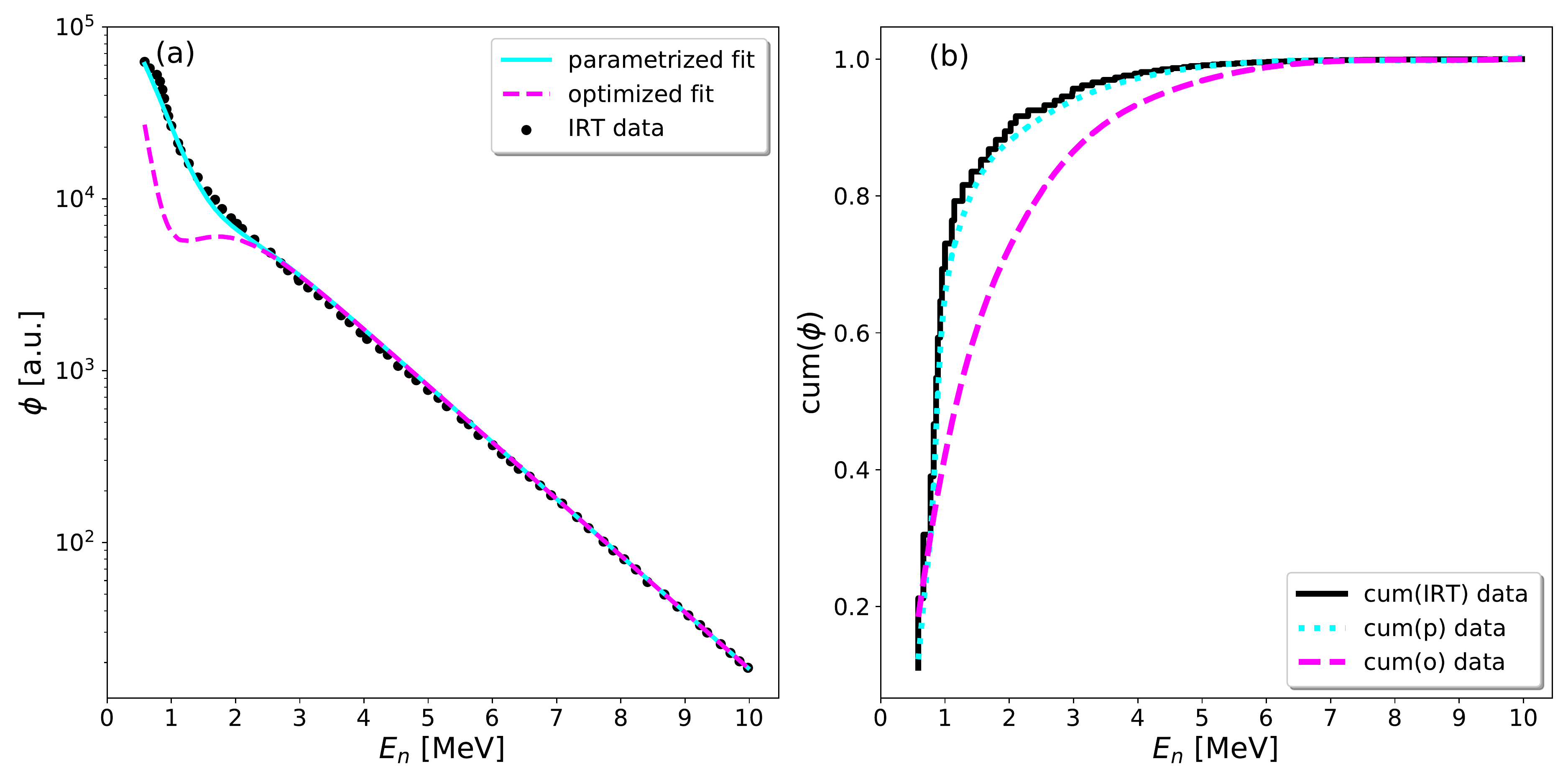}
  \caption{\label{fig:KS} (a) The solid cyan curve describes the overall fit to the sampled IRT-M data \cite{id:ATLAS} embodied by Eq.~(\ref{eq:phi_overall}).  The dashed magenta curve represents the optimized $kT$-adjusted version of this function.  (b) The black curve illustrates the normalized cumulative empirical distribution function (EDF) for the sampled IRT-M data, the cyan curve shows the corresponding parameterized continuous distribution function [CDF(p)] and the magenta curve represents that of the optimized continuous distribution function [CDF(o)].}
\end{figure*}

Given that we are working with only three data points and one adjustable parameter, there will be a small number of degrees of freedom (ndf): ${\rm ndf} = 3-1 = 2$.  In these circumstances and assuming Poisson statistics, the ideal $\chi^{2} \not =  1$; for ${\rm ndf} =2$ the reduced $\chi^{2}$, i.e., $\chi^{2}/{\rm ndf} \approx 0.35$ \cite{beringer:12}.  We find that a correlation coefficient $\rho_{ij} > 0.8$ cannot reproduce the expected $\chi^{2}/{\rm ndf}$ of 0.35.  For the off-diagonal matrix elements, values of $\rho_{ij}$ within the range $0 < \rho_{ij} \lesssim 0.75$ reproduce the expected $\chi^{2}/{\rm ndf}$ consistent with $0.125 \leq kT \leq 0.185$~MeV.  The range on $\rho_{ij}$ seems reasonable given that many correlated uncertainties cancel upon taking ratios to determine $B_{A}$ and $B_{kT}$.  Accordingly, adjustments to $kT$ demonstrate that $kT = 0.155(30)$~MeV satisfies the $\chi^{2}$ criterion.  Our corresponding best-fit flux-weighted angular-distribution-corrected $\gamma$-ray production cross sections for the three strongest $^{56}$Fe transitions are presented in Table~\ref{tab:fwa_opt}, revealing improved agreement to within $1\sigma$ in each case with the Atlas-reported measurements \cite{id:ATLAS}.

The optimized $kT$-adjusted fit is shown in Fig.~\ref{fig:KS}a in comparison to the original fit given by Eq.~(\ref{eq:phi_overall}) and parameterized according to the IRT-M reactor data \cite{id:ATLAS}.  Figure~\ref{fig:KS}b shows the corresponding normalized cumulative empirical distribution function (EDF) for the sampled IRT-M data, together with the normalized cumulative continuous distribution functions for both the parameterized [CDF(p)] and optimized [CDF(o)] fits to the data.  We can assess the difference between the sampled data and continuous distribution functions by performing a Kolmogorov-Smirnov (KS) test \cite{chakravarti:67}.  The KS statistic is given by the supremum between the sampled EDF and the hypothesized CDF
\begin{equation}
  D_{N} = {\rm sup}|\phi(E)_{{\rm CDF}(x)} - \phi(E)_{\rm EDF}|,
  \label{eq:ks}
\end{equation}
where $N$ is the number of data points sampled from the EDF, and $x$ refers to either the original parameterized (p) or the optimized (o) CDF.  In our test, $N=67$ and we find $D_{67} = 0.030$ for CDF(p), while $D_{67} = 0.236$ for CDF(o).  Critical $p$-values for different levels of significance ($\alpha$) and confidence levels (CL) are listed in Table~\ref{tab:p-vals}.  Here, the critical $p$-value is given by $p = 0.199$ at $\alpha = 1\%$ (${\rm CL}=99\%$), and $p = 0.131$ at $\alpha = 20\%$ (${\rm CL = 80\%}$).  Strictly speaking, the KS test only applies when the distribution is fixed beforehand, and therefore, it stands to reason that the KS statistic for the CDF(p) is substantially smaller than the critical $p$-value even at {\rm CL=80\%} because the derived CDF has already been fit to the empirical sample data.  However, the CDF(o) has been optimized in such a way that the KS test illustrates the degree of divergence from the original CDF(p), where it is found that the KS statistic for the CDF(o) is larger than the critical $p$-value even at {\rm CL=99\%}.  This suggests that the optimized $kT$-adjusted flux spectrum is significantly different to the neutron flux reported in the Baghdad Atlas \cite{id:ATLAS}.  Even though the parametrized fit to the Atlas-reported flux can reproduce the measured data to within $2\sigma$, further investigations should be carried out to help pin down an appropriate functional form of the neutron-flux spectrum to help provide a better description of the measured data.  An additional limitation of the current method is that the $^{56}$Fe data are only sensitive to neutron energies above the excitation threshold of 861.9~keV.

\begin{table}[h]
  \caption{ \label{tab:p-vals} Critical $p$-values for a distribution described by $N$ data points for $\alpha = 0.01$ (${\rm CL} = 0.99$) to $\alpha = 0.20$ (${\rm CL} = 0.80$).  The final row shows the corresponding $p$-values for an $N=67$ sample.}
  \begin{center}
    \begin{tabular}{c|ccccc} \hline\hline
      $\alpha$ & 0.01 & 0.05 & 0.10 & 0.15 & 0.20 \\
      CL & 0.99 & 0.95 & 0.90 & 0.85 & 0.80 \\ \hline
      $p$ & $\frac{1.63}{\sqrt{N}}$ & $\frac{1.36}{\sqrt{N}}$ & $\frac{1.22}{\sqrt{N}}$ & $\frac{1.14}{\sqrt{N}}$ & $\frac{1.07}{\sqrt{N}}$ \\
      $p(N=67)$  & 0.199 & 0.166 & 0.149 & 0.139 & 0.131 \\
      \hline\hline
    \end{tabular}
  \end{center}
\end{table}

As a final consideration, we can also compare our $kT$-optimized flux-weighted cross sections to those from the recent $^{56}$Fe($n,n'\gamma$) measurement using the Fast Neutron-induced Gamma Spectrometry (FaNGaS) setup at the FRM-II (Forschungs-Neutronenquelle Heinz-Maier-Leibnitz) research reactor by Ilic \cite{ilic:20}, where they have also determined average flux-weighted angular-distribution-corrected cross sections for the 846.8-, 1238.3- and 1810.8-keV $\gamma$ rays; presented in Table~\ref{tab:fwa_opt} for comparison.  Although our results are significantly smaller, Table~\ref{tab:fwa_opt} shows that all results are in proportion.  The absolute differences, however, are not surprising given a much higher average neutron energy of 2.3~MeV in the FRM-II measurement \cite{ilic:20}, while our $kT$-optimized flux-weighted neutron expectation energy is only 0.631~MeV according to Eq.~(\ref{eq:wexpt5}), whereupon we replace the lower-limit $E_{\rm th}$ with $E=0$.  Interestingly, upon scaling our cross sections by the ratio of neutron expectation energies
\begin{equation}
  \langle \sigma_{\gamma_{i}} \rangle_{S} = \langle \sigma_{\gamma_{i}} \rangle_{W} \frac{\langle E_{n}^{\rm FRM} \rangle}{\langle E_{n} \rangle},
  \label{eq:frm}
\end{equation}
where $\langle E_{n}^{\rm FRM} \rangle$ denotes the FRM-II neutron expectation energy \cite{ilic:20}, we find that our scaled cross sections $\langle \sigma_{\gamma_{i}} \rangle_{S}$ are in agreement with those from the FRM-II, i.e., $\langle \sigma_{\gamma_{i}} \rangle_{S} \approx \langle \sigma_{\gamma_{i}} \rangle_{\rm FRM}$ as shown in Table~\ref{tab:fwa_opt}.

\section{Database scope and structure}

The data in the Baghdad Atlas is valuable for many applications ranging from nuclear non-proliferation, active neutron-interrogation studies and benchmarking nuclear-reaction models in the fast-fission neutron-energy region.  However, its use is limited by the fact that the data was only available in printed form \cite{id:ATLAS}.  To enhance its utility we have compiled the data into a set of CSV-style ASCII tables and developed software to produce a locally-installed SQLite relational database. This allows for far greater dissemination, increasing the database accessibility for the international community. The complete package may be downloaded \cite{id:BNDG,id:NNDC} and contains the following components:

\begin{itemize}
\item Scripts to compile the data into a SQLite database for both Python 2 and Python 3.  The {\tt Makefile} provided will test for the appropriate version and run with the necessary settings.
\item Open-source C-code to produce the shared-object dynamic extension-functions library allowing for enhanced functionality in SQL transactions that are not part of the standard SQLite3 library. This library will then provide the user with access to common mathematical (e.g., {\tt cos}, {\tt sin}, {\tt tan}, {\tt exp}, {\tt log}, {\tt log10}, {\tt sqrt}, {\tt pi}, etc.), string (e.g., {\tt replicate}, {\tt replace}, {\tt reverse}, etc.), and aggregate (e.g., {\tt variance}, {\tt mode}, {\tt median}, etc.) functions in SQL queries using the OS libraries or provided definitions. The {\tt Makefile} will establish the correct OS-kernel name and compile the library accordingly.
\item The complete set of CSV-style ASCII data files compiled for 76 different elements in the range $3 \leq Z \leq 92$.  This includes data sets from 76 natural samples and 29 isotopically-enriched samples (105 data sets in total).
\item A Jupyter Notebook illustrating Python-based methods for interacting with and visualizing the data.
\item Jupyter Notebooks to reproduce the flux analysis reported in this paper in addition to the angular distribution analysis.  An angular momentum coupling coefficient calculator is also provided therein.
\item Several SQL scripts based on standard SQL-transaction methods exemplifying database interaction.  The compiled OS-dependent extension-functions library will be initialized as appropriate where required during the build process.
\item A PDF of the Baghdad Atlas \cite{id:ATLAS} provided for reference.
\item An HTML manual describing the software-installation procedure and data-retrieval methods.  An overview of the Baghdad Atlas data is also provided with this documentation.  This HTML manual is also available online \cite{id:BNDG}.
\end{itemize}

\subsection{Database schema}

The data structure for the ($n,n'\gamma$) data is described using two relational tables in an SQL schema: {\tt nucleus} and {\tt sample}.  The {\tt nucleus} table represents the nuclear-structure type class of information including: chemical symbol and atomic number along with flags to indicate element or enriched isotope identification, $\gamma$-ray energies and intensities together with their associated uncertainties, excitation energies of the states populated, and residual-nucleus reaction products.  Associated properties of the aforementioned quantities are also contained in this table.  The {\tt sample} table contains all meta-data associated with the irradiated sample including: element/isotope identification properties, atomic number, mass number, irradiation exposure period, chemical composition, mass and enrichment data.  Normalization $\gamma$-ray properties (energies, scaling factors and uncertainties) are also listed in this table.  A detailed explanation of the schema is available in the supporting documentation \cite{id:BNDG}.

\begin{figure}[t]
  \includegraphics[angle=0,width=\linewidth]{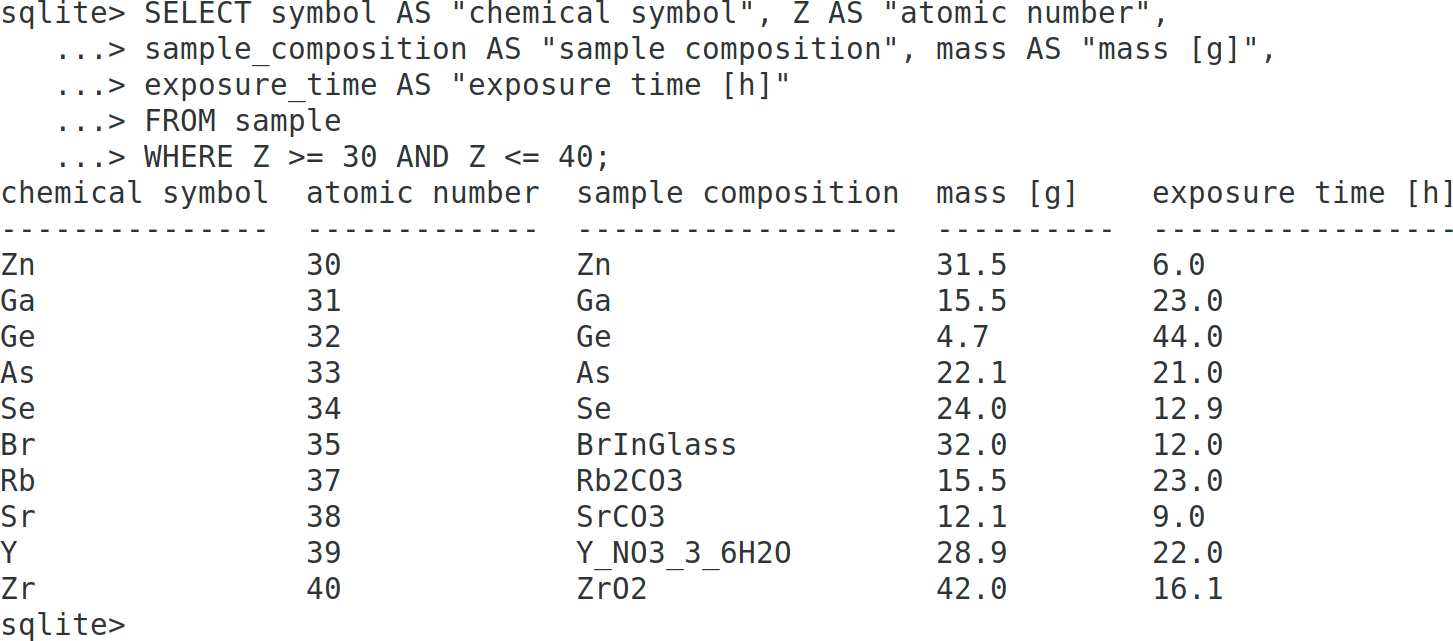}
  \caption{\label{fig:3} Example of a SQLite transaction using conditional arguments to access information in the Baghdad Atlas database.  See text for details.}
\end{figure}

\subsection{Additional requirements}

The software provided with this package is intended to create a SQLite database on Linux and Mac OS X platforms.  Additional system requirements needed to build the database and run this software include: the SQLite3 database engine \cite{id:SQLite}; the GNU C compiler \cite{id:GNUC}; Python 2.7 or Python 3 \cite{id:Python}.  In addition, users that choose to run the provided Jupyter Notebook ``as is'' will require installation of several third-party Python libraries.  These requirements along with other recommendations are outlined in the supporting online documentation \cite{id:BNDG}.

\section{Database-retrieval methods}

Since we provide access to the source data itself, together with descriptive documentation, this allows users the freedom to parse and manipulate the data sets according to individual needs.  However, the database utility and associated software described in this paper offers users a readily-accessible and convenient means for interacting with and visualizing the data.  Once built and installed to the appropriate location, database queries can be performed by invoking transactions based on the SQLite syntax.  This provides users with a variety of options for retrieving customized data sets based on conditional arguments.

\subsection{Terminal command line}

The SQLite engine \cite{id:SQLite} is an embedded SQL database engine providing a terminal-based frontend to the SQLite library that is able to evaluate queries interactively and display the corresponding results.  This method is particularly useful for rapid evaluation of simple queries. Figure~\ref{fig:3} illustrates this concept using the SQLite ``interpreter'' to query the Baghdad Atlas database and retrieve \textit{chemical symbol, atomic number, sample composition, mass,} and \textit{exposure time} information for selective entries from the {\tt sample} relational table that satisfy the atomic number condition: $30 \leq Z \leq 40$.  Alternatively, more complicated queries are better handled by reading in scripts, either through the interpreter or, because the SQLite engine also supports batch-mode processing, directly via the command line.  Additionally, for users that prefer to interact with the database through a graphical user interface, a variety of open-source cross-platform distributions are available, e.g., Refs.~\cite{id:DBBrowser}, \cite{id:Studio}.  Several SQL scripts illustrating database interaction and suitable for processing using the aforementioned methods are bundled with the software package.

\begin{figure}[t]
  \includegraphics[angle=0,width=\linewidth]{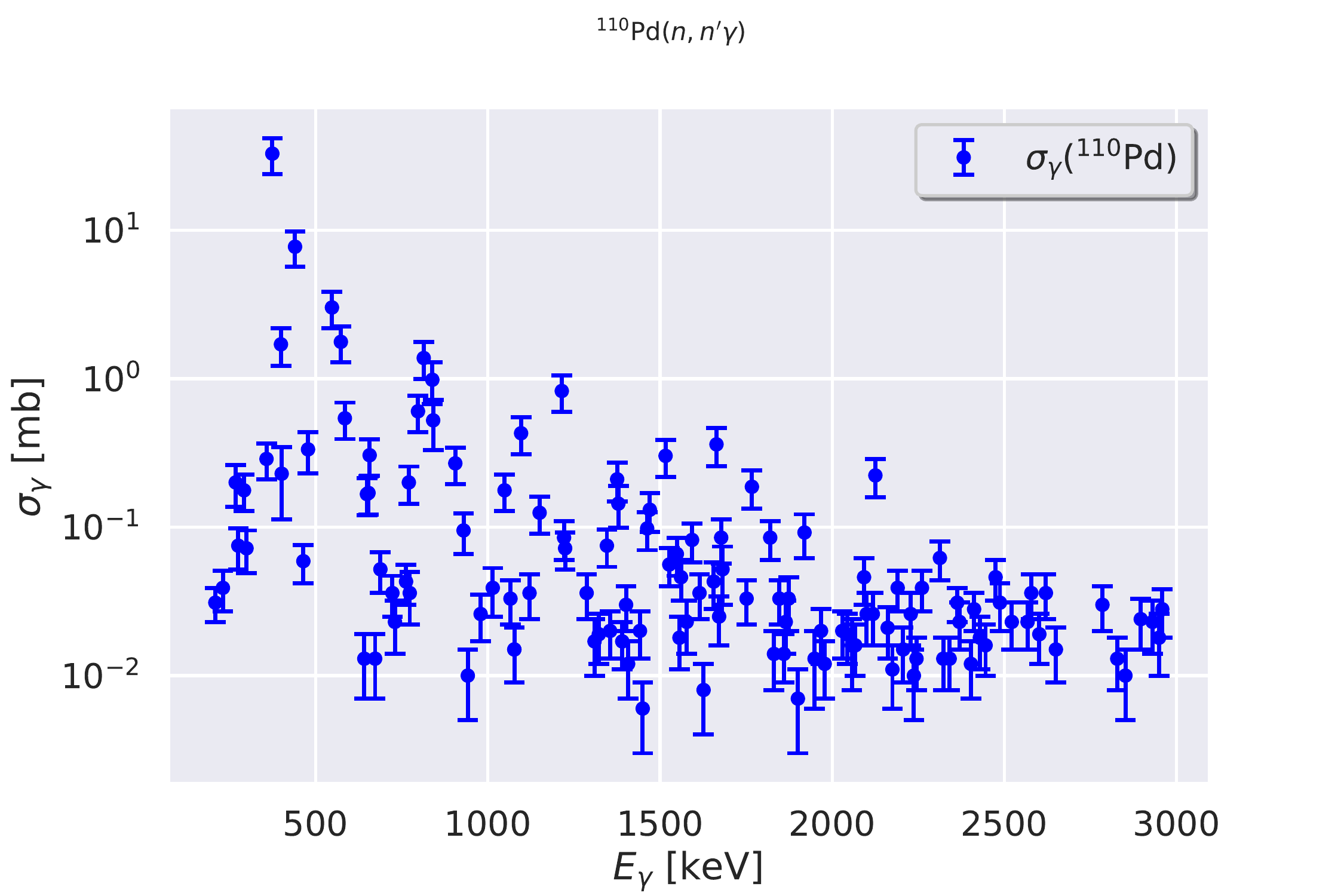}
  \caption{\label{fig:4} Absolute partial $\gamma$-ray cross sections for the $^{110}$Pd($n,n'\gamma$) reaction extracted from the Baghdad Atlas database using the Python Notebook automation methods.}
\end{figure}

\begin{figure}[t]
  \includegraphics[angle=0,width=\linewidth]{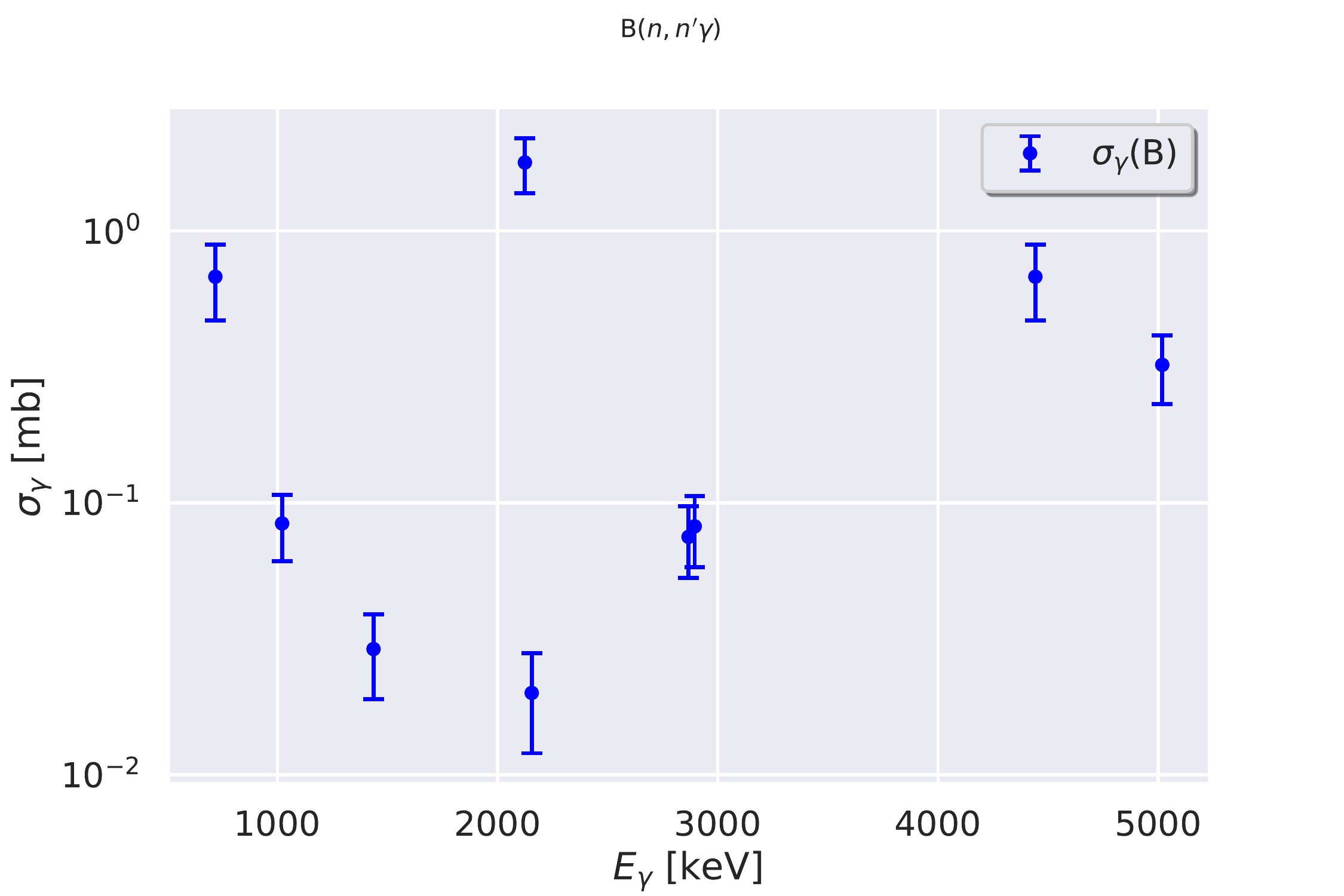}
  \caption{\label{fig:5} Absolute partial $\gamma$-ray cross sections for the $^{\rm nat}$B($n,n'\gamma$) reaction extracted from the Baghdad Atlas database using the Python Notebook automation methods.}
\end{figure}

\subsection{Jupyter Notebook}

The Jupyter Notebook ~\cite{id:Jupyter} provides users with a browser-based frontend interface to interact with the SQLite libraries. One of the main advantages of creating Jupyter Notebook projects is that it greatly enhances and facilitates project sharing among researchers, allowing users to create collaborative and reproducible narratives. Although this method requires additional Python-code overhead, it provides a convenient means for populating arrays and lists of data that can be sorted and filtered according to specific requirements.  Also, the Jupyter Notebook provides access to a wide range of other useful libraries allowing for implementation of high-level methods to augment the analysis of the data, as well as a high degree of flexibility in displaying the results$-$including inline visualization of the data for on-the-fly inspection.

Many nuclear data applications require information regarding absolute partial $\gamma$-ray cross sections.  In recognition of this need, the Notebook bundled with this software package has been developed with automation processes to allow users to easily generate tables of cross sections (written to file) as well as presentation-style plots according to a user-defined nucleus: \textit{atomic mass, atomic number,} and \textit{chemical symbol}.  For example, in the case of the enriched isotope $^{110}$Pd, we simply define: {\tt Z = int(46)}, {\tt A = int(110)}, and {\tt Chem\_symb = str(``110Pd'')} in the appropriate cell of the Notebook prior to execution of the cell to generate the plot of energy-dependent partial $\gamma$-ray cross sections in Fig.~\ref{fig:4}.  For natural samples of elemental composition, the atomic mass is defined as $A = 0$.  Thus, in the case of a natural boron sample, for example, we define: {\tt Z = int(5)}, {\tt A = int(0)}, and {\tt Chem\_symb = str(``B'')} to generate the plot shown in Fig.~\ref{fig:5}.  The {\tt CrossSection} class implemented in the Notebook that is used to generate these normalized data sets contains constants associated with the deduced flux-weighted average cross section for $^{56}$Fe, $\langle \sigma_{\gamma 847} \rangle = 143(29)$~mb (see Sect.~\ref{sect:3}).  Adjustment of these class constants also provides users with the freedom to renormalize data sets according to different expectation values obtained from other neutron-data libraries or experimental measurements in a straightforward manner.  Moreover, revisions to the adopted normalization cross section do not impact the source branching-ratio data stored on disk.  It should be noted that the scaled cross sections described here are determined in accordance with Eq.~(\ref{eq:scale1}).  However, because the angular-distribution effect can pose a significant correction, if sufficient information is available then Eq.~(\ref{eq:scale2}) provides a more appropriate scaling method.

\section{Summary and outlook}

This first release of the Baghdad Atlas database is designed to serve the needs of the applications community.  The energy levels, $\gamma$-ray energies and intensities are those stated in the Atlas \cite{id:ATLAS} itself and have not, in general, been reconciled to match the adopted values in the ENSDF database \cite{id:ENSDF}, although up to date structure information has been used to identify $\gamma$-ray doublets (reported as unresolved doublets for now).  Our intention is to issue periodic revisions where the $\gamma$-ray and level-scheme information for specific nuclei has been updated to match modern values in ENSDF \cite{id:ENSDF}, that can subsequently be used for more general dissemination to the wider-user community.

In addition to its immediate use to the applications community, the Baghdad Atlas can serve as a valuable resource to both the nuclear structure and reactions evaluations communities. Since ($n,n'$) at fast-reactor neutron energies proceeds via both direct and compound processes it provides a non-selective insight into the properties of off-yrast levels over a far wider range of spins than thermal and epithermal neutron capture.  This property of ($n,n'$) has made it an attractive area of study to many research groups in the international nuclear science community. This is exemplified by the plethora of papers on $^{56}$Fe($n,n'\gamma$) by groups from Russia \cite{demidov:04}, Los Alamos \cite{fotiades:10,fotiades:04}, GELINA \cite{negret:14}, Dresden \cite{beyer:14}, and FRM-II \cite{ilic:20}.

It should be noted, however, that our attempts to model the flux with a simple Maxwellian or a modified Watt spectrum fail to reproduce the neutron-flux spectrum reported in the Baghdad Atlas \cite{id:ATLAS}.  Only by developing a compound function incorporating these models in distinct regions of the spectrum were we able to reproduce the reported flux.  Furthermore, our fitted parametrizations differ significantly to the corresponding expectation values for a $^{235}$U fission spectrum \cite{fluence:IAEA}.  Accordingly, our interpretation of the flux represents a significant source of uncertainty, and perhaps imprecision, in the determination of the flux-weighted cross sections.  However, the treatment presented in this paper is largely for pedagogical purposes and it is our hope that the adopted procedure can also be employed with better-understood neutron fluxes.  In that respect, we aim to continue our investigations into the nature of an appropriate neutron flux to describe accurately the valuable data presented in the Baghdad Atlas.

Lastly, we hope that this SQLite database will provide a useful tool for the reactions evaluation community by providing easy access to energy-integrated data to aid in the benchmarking process needed for the validation of evaluated neutron-data libraries including ENDF/B-VIII.0 \cite{brown:18}, the Japanese Evaluated Nuclear Data Library (JENDL-4.0) \cite{id:JENDL}, the Joint Evaluated Fission and Fusion File (JEFF-3.3) \cite{id:JEFF}, and the TALYS-generated Evaluated Nuclear Data Library (TENDL) \cite{id:TENDL}.  This use of the database has already been demonstrated since the data were recently compiled into the EXFOR \cite{id:EXFOR, zerkin:18} library.

\section*{Acknowledgments}

This material is based upon work supported by the Department of Energy National Nuclear Security Administration under Awards No. {DE-NA0003180} and No. {DE-NA0000979} at the University of California, Berkeley.  This work is also performed under the auspices of the National Nuclear Security Administration of the US Department of Energy at the Los Alamos National Laboratory under Contract No. 89233218CNA000001, and is supported in part by the Lawrence Berkeley National Laboratory under Contract No. DE-AC02-05CH11231 for the US Nuclear Data Program.  The authors would like to thank Dr.~Andrej Trkov for {\sl ``saving''} the Baghdad Atlas, in addition to Dr.~Rick Firestone, Dr.~Bradley Sleaford, and Dr.~David Brown for helpful discussions and feedback.  One of the authors, Aaron, wishes to thank Mrs. Geraldine Norma Hurst, for everything.

\bibliography{refs_nng}

\end{document}